\documentclass[iop, revtex4]{emulateapj}
\usepackage{amsmath}
\usepackage{natbib}
\usepackage{booktabs}
\usepackage{hyperref}
\usepackage{xcolor}
\usepackage{color}
\usepackage{epsfig}

\shorttitle{Clustering of Lyman-alpha Emitters Around Quasars at $z\sim4$}
\shortauthors{Garc\'ia-Vergara, C. et al.}

\begin{document}

\title{Clustering of Lyman-alpha Emitters Around Quasars at $z\sim4$\footnotemark[1]}\footnotetext[1]{Based on observations collected at the European Organization for Astronomical Research in the Southern Hemisphere, Chile. Data obtained from the ESO Archive, Normal program, service mode. Program ID: 094.A-0900.}

\author{Cristina Garc\'ia-Vergara \altaffilmark{2, 3, 4}}
\author{Joseph F. Hennawi \altaffilmark{4,5}}
\author{L. Felipe Barrientos \altaffilmark{3}}
\author{Fabrizio Arrigoni Battaia \altaffilmark{6}} 

\altaffiltext{2}{Leiden Observatory, Leiden University, P.O. Box 9513, 2300 RA Leiden, The Netherlands.}
\altaffiltext{3}{Instituto de Astrof\'isica, Pontificia Universidad Cat\'olica de Chile, Avenida Vicu\~na Mackenna 4860, Santiago, Chile.}
\altaffiltext{4}{Max-Planck Institut f\"ur Astronomie (MPIA), K\"onigstuhl 17, D-69117 Heidelberg, Germany.}
\altaffiltext{5}{Department of Physics, University of California, Santa Barbara, CA 93106, USA}
\altaffiltext{6}{Max-Planck Institut f\"ur Astrophysik (MPA), Karl-Schwarzschild-Str. 1, 85741 Garching bei M\"unchen, Germany}
\email{garcia@strw.leidenuniv.nl}

\begin{abstract}
The strong observed clustering of $z>3.5$ quasars
indicates they are 
hosted by massive ($M_{\rm{halo}}\gtrsim10^{12}\,h^{-1}\,\rm{M_{\odot}}$) dark matter halos. 
Assuming  quasars and galaxies trace the same large-scale structures, 
this should also manifest as strong clustering of galaxies around quasars.
Previous work on high-redshift quasar environments,
mostly focused at $z>5$, have failed to find convincing evidence for
these overdensities. Here we conduct a survey for Lyman alpha
emitters (LAEs) in the environs of 17 quasars at $z\sim4$ probing
scales of $R\lesssim7\,h^{-1}\,{\rm{Mpc}}$. We measure
an average LAE overdensity around quasars of 1.4 for our full sample,
which we quantify by fitting the quasar-LAE cross-correlation function.
We find  consistency with a power-law shape with correlation length of
$r^{QG}_{0}=2.78^{+1.16}_{-1.05}\,h^{-1}\,{\rm{cMpc}}$ for a fixed
slope of $\gamma=1.8$. We also measure the LAE
auto-correlation length and find
$r^{GG}_{0}=9.12^{+1.32}_{-1.31}\,h^{-1}$\,cMpc ($\gamma=1.8$), which is $3.3$ times
higher than the value measured in blank fields. Taken
together our results clearly indicate that LAEs are significantly clustered around $z\sim4$ quasars.
We compare the observed clustering with the expectation from a deterministic bias model, whereby LAEs and quasars probe
the same underlying dark matter overdensities, and find that our measurements
fall short of the predicted overdensities by a factor of 2.1.
We discuss possible explanations for this discrepancy including large-scale
quenching or the presence of excess dust in galaxies near quasars.
Finally, the large cosmic variance from field-to-field observed in our
sample  (10/17 fields are actually underdense) cautions one
from over-interpreting studies of $z\sim6$ quasar
environments based on a single or handful of quasar fields. 
\end{abstract}

\section{Introduction}
\label{sec:intro}

Understanding large-scale structure at early cosmic epochs, and
how high-redshift $z\gtrsim 4$ quasars and galaxies fit into the
structure
formation hierarchy, remains an important challenge for observational astronomy.
In the current picture, every massive galaxy resides in a massive dark matter halo and is thought
to have gone through a luminous quasar phase fueling the growth of a central supermassive black hole.
If the strong correlations between black hole mass and galaxy mass observed locally \citep{Magorrian98, Ferrarese00, Gebhardt00} were also
in place at early times, then one expects bright quasars at $z\gtrsim 4$ to be signposts for
massive structures in the young Universe.
Many 
theoretical studies have sought to understand quasar
evolution \citep[e.g.][]{springel05,angulo12}
in a
cosmological context, and generically predict that high-redshift
$z\gtrsim 4$ quasars should reside in massive halos $M_{\rm halo}\gtrsim
10^{12}\,h^{-1}\, \rm M_{\odot}$ , and thus trace massive protoclusters. Such protoclusters can be identified as large overdensities
of galaxies, which are expected to be detectable on typical scales of $R\sim4.5 - 9$  Mpc at $z\sim4$ \citep{Chiang13}. However, observational constraints have yet to
definitively confirm this hypothesis, primarily due to the challenges
of detecting galaxies at such high redshifts. 

A fundamental result of the $\Lambda$CDM structure formation paradigm
is that the clustering of a population of objects can be directly
related to their host dark halo masses \citep{Mo96}.  The
auto-correlation length $r_0$ of quasars is observed to rise steadily
 over the range $0 \lesssim z\lesssim 2$,
\citep{Porciani06,Myers06, Shen07,White12,Eftekharzadeh15}, and then dramatically
steepens from $z\sim 2$ to $z\sim 4$, the highest redshift for which
the auto-correlation has been measured.  \citet{Shen07} measured the
quasar auto-correlation using a sample of $\sim 1,500$ quasars at $z\geq3.5$
from the Sloan Digital Sky Survey (SDSS; \citealt{York00}), and
inferred a correlation length of $r_0=24.3\,h^{-1}\,{\rm Mpc}$ (for a
fixed slope for the correlation function of $\gamma=2.0$). This
measurement indicates that $z\geq3.5$ quasars are the most highly
clustered population at this epoch, implying very massive ($M_{\rm
  halo}>(4-6) \times 10^{12}\,h^{-1}\,\rm M_{\odot}$) host dark matter
halos \citep[see also][]{White08,Shen09}.

An alternative and complementary approach is to directly characterize
the density field around quasars by searching for galaxies clustered
around them. But to date the results of such studies are
difficult to interpret and it remains unclear whether $z\gtrsim 4$ quasars reside in overdensities
of galaxies,  as should be implied by the strong auto-correlation measured by \citet{Shen07}.
Currently, about 30 quasar fields
have been studied for this purpose at $z\geqslant4$, and some of them
show overdensity of galaxies \citep{Stiavelli05, Zheng06, Kashikawa07,
  Kim09, Utsumi10, Capak11, Swinbank12, Morselli14, Adams15,
  Balmaverde17, Kikuta17, Ota18}, whereas others exhibit a similar number
density of galaxies compared with blank fields \citep{Willott05,
  Kim09, Banados13, Husband13, Simpson14, Mazzucchelli17, Goto17,
  Ota18}. \citet{Uchiyama18} performed a more systematic search for
associations of 171 SDSS quasars at $z\sim4$ with Lyman break galaxies
(LBGs) overdensities detected in a $\sim121$ square degrees survey,
and find that on average quasars are not associated with  overdensities.

Several different factors could explain these contradictory
results. First, nearly all of the aforementioned studies aim to detect
overdensities of galaxies around individual or at most a handful of
quasars, and the large statistical fluctuations expected from cosmic
variance could explain why they have been inconclusive.  Second,
comparing different studies is complicated by the fact that they often
select different types of galaxies. Many searches have been conducted
for LBGs around quasars, but the wide redshift range $\Delta z \sim 1$
resulting from photometric selection
\citep[e.g.][]{Ouchi04a,Bouwens07,Bouwens10}, make them particularly
susceptible to projection effects that will dilute any clustering
signal. Other work targeting Lyman alpha emitters (LAEs) over a much
narrower redshift range of $\Delta z \sim 0.1$ should result in less
dilution, but the fact that LAEs are intrinsically less strongly
clustered than LBGs \citep[e.g.][]{Ouchi04b, Kashikawa06, Ouchi10, Ouchi18}, complicates the
comparison. Finally, past work has quantified clustering via different
statistics, on different physical scales (often limited by the
instrument field-of-view), and may suffer from systematic errors in
the determination of the expected background number density of
galaxies, which is a crucial issue given that clustering can be
strongly diluted by projection effects. In summary, the diverse and
contradictory results emerging from studies of quasar environs at
$z\gtrsim 4$ may simply reflect a lack of sensitivity and heterogeneous
analysis approaches. 

One strategy for overcoming these complications is to focus on measuring the
quasar-galaxy cross-correlation function, and target a large sample of quasars 
leveraging the statistical power to increase sensitivity. This cross-correlation function
has several advantages: i) it quantifies not only the over/under density
of galaxies but also their radial distribution about the quasar, ii) it can
be easily related to the respective auto-correlations of the quasar and galaxy
samples, and iii) similar to the auto-correlation, it provides an independent
method to estimate quasar host halos masses.
The quasar-galaxy
cross-correlation function has been measured by many studies at $z<3$
\citep[e.g.][]{Adelberger05,Padmanabhan09, Coil07, Trainor12, Shen13},
but only few works have attempted a measurement at higher redshifts. 
\citet{Ikeda15} measured the angular quasar-LBG
cross-correlation function using a sample of 16 spectroscopically
confirmed quasars at $3.1<z<4.5$. They report an upper limit for the
cross-correlation length of $r_{0}< 10.72\,h^{-1}\,{\rm Mpc}$ (for a
fixed slope of $\gamma=1.8$). Similarly, \citet{He18} used spectroscopic SDSS
luminous quasars at $3.4<z<4.6$ to cross-correlate with LBGs, reporting a cross-correlation length of
$r_{0}=5.43^{+2.52}_{-5.06}\,h^{-1}\,{\rm Mpc}$ (for a fixed slope of
$\gamma=1.86$).  \citet{garciavergara17} measured the projected
quasar-LBG cross-correlation function on scales of $0.1\lesssim
R\lesssim9\,h^{-1}\,{\rm Mpc}$ by using a sample of six spectroscopic
quasars at $z\sim4$. They find that LBGs are strongly clustered around
quasars\footnote{Note that in this work, LBGs were selected using a
  novel NB technique that ensures a redshift coverage of $\Delta z
  \sim 0.3$ and then this result is not affected by typical LBG
  redshift uncertainties that could be enhancing the real LBG number
  counts.}, with a cross-correlation length of
$r^{QG}_{0}=8.83^{+1.39}_{-1.51}\,h^{-1}\,{\rm Mpc}$ (for a fixed
slope of $\gamma=2.0$), which agrees with the expected
cross-correlation length computed using the LBG auto-correlation
length and the quasar auto-correlation length assuming a deterministic
bias model

In this paper, we present the first measurement of the $z\sim4$ quasar-LAE cross-correlation function based on narrow- and broad-band observations performed using VLT/FORS2 on a sample of 17 quasar fields. The spectroscopic quasars have been selected to have small redshift uncertainties such that
their Ly$\alpha$ emission line lands in the central part of the narrow band (NB) filter employed. There are
several motivations for performing this study at $z\sim4$. First, the auto-correlation function of both quasars \citep{Shen07} and LAEs are
\citep{Ouchi10} are well measured by previous work, which allow us to compute the expected strength of the cross-correlation assuming quasars and galaxies trace the same underlying dark matter
overdensities. Second, at $z\sim4$ the LAE luminosity function is well measured \citep[e.g.][]{Ouchi08,Cassata11,Drake17,Sobral18}, and therefore we can use it to compute the expected LAE number density in blank fields, a key ingredient in a cross-correlation measurement. Third, this luminosity function implies that
$z\sim 4$ LAEs are sufficiently numerous and bright that one ought to detect the expected level of the cross-correlation given a reasonable investment of
telescope time. Finally, an independent measurement of the quasar-LBG cross-correlation exists at nearly the same redshift
\citep{garciavergara17}, enabling a comparison of the clustering of two different galaxy populations around quasars at the same cosmic epoch.

This paper is organized as follows. In section \S~\ref{sec:data} we provide information on the sample and describe the observations and data reduction. In section \S~\ref{sec:number_density} we discuss the selection of the LAEs in quasar fields and compute the number density of LAEs in our sample. The clustering analysis is presented in section \S~\ref{sec:clustering}, and the implications of our results are discussed in section \S~\ref{sec:discussion}. We summarize the work in section \S~\ref{sec:sum}.

Throughout this paper magnitudes are given in the AB system
\citep{Oke74, Fukugita95} and we adopt a cosmology with $H_{0}=100\,
h\,{\rm km\,s}^{-1}\,{\rm Mpc}^{-1}$, $\Omega_{m}=0.30$ and
$\Omega_{\Lambda}=0.70$ which is consistent with \citet{Planck18}. Comoving and proper Mpc are denoted as ``cMpc" and ``pMpc",
respectively.

\section{Observations and Data Reduction} 
\label{sec:data}

\subsection{Quasar Selection} 
\label{ssec:targets}

\begin{figure}
\centering{\epsfig{file=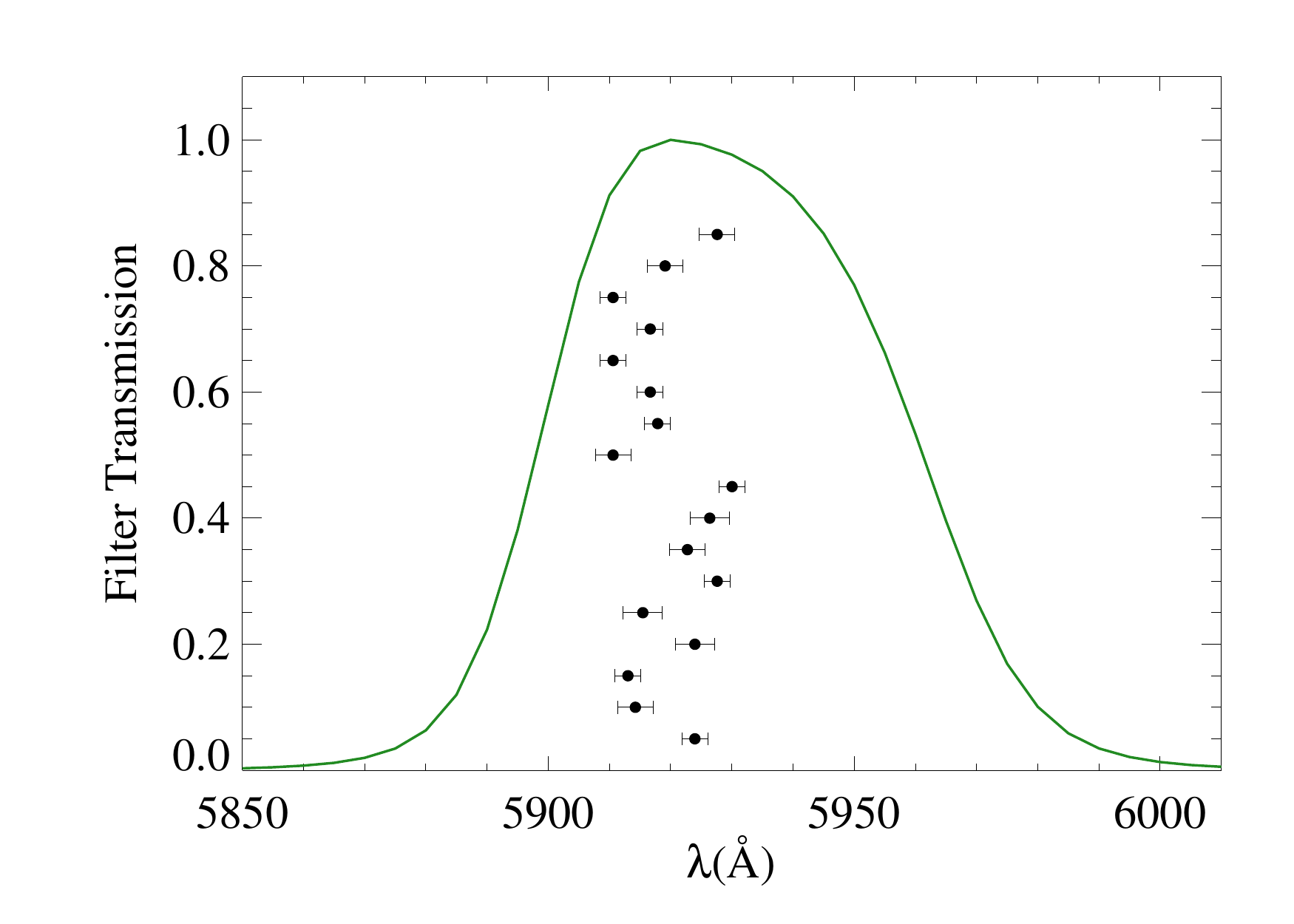, width=\columnwidth}}
\caption{\label{fig:qso_z} Normalized NB filter transmission curve (green). Each point shows the wavelength in which the Ly$\alpha$ emission line of each quasar lie at their systemic redshift reported in Table ~\ref{table:qso_prop}. Y-axis position of the points is arbitrary. Quasar redshifts have been determined from the quasar SDSS/BOSS spectra by centroiding one or more of their rest-frame UV emission lines, SIV, CIV and CIII]. Then, we determined systemic redshifts by using the calibration of emission line shifts from \citet{Shen07} (see details in \S~\ref{ssec:targets}). The Ly$\alpha$ emission line for all the quasars including their 1$\sigma$ errors lies in the central part of our NB filter.\\}
\end{figure}

An important requirement to perform clustering analysis of galaxies around quasars is to reduce the impact of projection effects which dilute the clustering signal that one
intends to measure. To ensure that LAEs are selected over a narrow redshift window, we designed an observing program  using a NB filter $\rm HeI/2500+54$ (hereafter HeI), centered at $\lambda_{\rm c} = 5930$\AA, with $\rm FWHM = 63$\AA\, (or equivalently $\rm 3,197\,km\,s^{-1}$ corresponding to a comoving distance of $26.2\,h^{-1}$\,cMpc). This filter was chosen to select LAEs at $z=3.88$. We have then selected 17 quasars from the SDSS and the Baryon Oscillation Spectroscopic Survey (BOSS; \citealt{Eisenstein11,Dawson13}) quasar catalog \citep{Paris14} to lie within a redshift window set by the NB filter. Specifically, we selected only quasars whose systemic redshifts
lie within the central $\rm 1,066\,km\,s^{-1}$ of the NB filter (i.e., $\rm 1/3 \times FWHM$),  corresponding to $\Delta z \sim 0.02$ in redshift space. It is well known that quasar systemic redshifts differ from redshifts determined from rest-frame UV emission lines because of outflowing/inflowing material in the broad-line regions of quasars \citep{Gaskell82, Tytler92, VandenBerk01, Richards02b, Shen07, Shen16, Coatman17}. We thus determined quasar redshifts taking their SDSS/BOSS spectra and centroiding one or more of their rest-frame
UV emission lines (SIV $\lambda$1396\AA, CIV $\lambda$1549\AA\, and CIII] $\lambda$1908\AA) using a custom line-centering code \citep{hennawi06b} and using the calibration of emission line shifts from \citet{Shen07} to estimate the systemic redshifts. By using this technique, the resulting $1\sigma$ redshift uncertainties are $\rm 521\,km\,s^{-1}$ ($\Delta z = 0.008$) when the set of lines SIV, CIV and CIII] is used, $\rm 714\,km\,s^{-1}$ ($\Delta z = 0.012$) when both the SIV and CIV lines are used, and $\rm 794\,km\,s^{-1}$ ($\Delta z = 0.013$) when only the CIV line is used. All these uncertainties in the quasar redshifts are much narrower than the NB filter width used in this work implying that our NB filter actually selects LAEs at the same redshift of the central quasar (see Fig.~\ref{fig:qso_z}). 

Since radio loud quasars are known to reside in dense environments
\citep[e.g.][]{Venemans07}, and are observed to cluster more strongly
\citep[e.g.][]{Shen09}, we chose to focus our study on radio-quiet
quasars. We thus discarded all quasars from our initial selection which
had a radio emission counterpart at 20cm reported in the the Faint
Images of the Radio Sky at Twenty-centimeters
\citep[FIRST;][]{Becker95} catalog.  Finally, we also discarded quasars
with high galactic extinctions ($A_{\lambda}>0.2$) toward their line
of sight, and with close bright stars that could affect the imaging. A
sample of 29 quasars satisfied all the mentioned criteria, and we
selected the final sample of 17 quasars based on their visibility
conditions and gave a higher priority to the brightest objects. A
summary of the properties of the quasars we targeted is given in
Table~\ref{table:qso_prop}.

\begin{deluxetable*}{l r r c r}
\tabletypesize{\scriptsize}
\tablecaption{Targeted quasar properties.\label{table:qso_prop}}
\tablewidth{0.8\textwidth}
\tablehead{
\colhead{Field}&
\colhead{RA (J2000)}&
\colhead{DEC (J2000)}&
\colhead{Systemic Redshift}&
\colhead{$i$}
}
\startdata
SDSSJ0040+1706 &    00:40:17.426 &  +17:06:19.78 &   3.873  $\pm$ 0.008&  18.91   \\
SDSSJ0042--1020 &    00:42:19.748 &  -10:20:09.53 &  3.865    $\pm$ 0.012&  18.57  \\
SDSSJ0047+0423 &    00:47:30.356 &  +04:23:04.73 &   3.864  $\pm$ 0.008&  19.94  \\
SDSSJ0119--0342 &    01:19:59.553 &  -03:42:16.51 &  3.873    $\pm$ 0.013&  20.49  \\
SDSSJ0149--0552 &    01:49:06.960 &  -05:52:18.85 &  3.866    $\pm$ 0.013&  19.80  \\
SDSSJ0202--0650 &    02:02:53.765 &  -06:50:44.54 &  3.876    $\pm$ 0.008&   20.64 \\
SDSSJ0240+0357 &    02:40:33.804 &  +03:57:01.59 &    3.872 $\pm$ 0.012&   20.03 \\
SDSSJ0850+0629 &     08:50:13.457 &	+06:29:46.91   & 3.875 $\pm$ 0.013 & 20.40 \\
SDSSJ1026+0329 &   10:26:32.976  & +03:29:50.63  &   3.878  $\pm$ 0.008&  19.74  \\
SDSSJ1044+0950 &   10:44:27.798  & +09:50:47.98  &   3.862  $\pm$ 0.012&  20.52  \\
SDSSJ1138+1303 &   11:38:05.242  & +13:03:32.61  &   3.868  $\pm$ 0.008&  19.10  \\
SDSSJ1205+0143 &   12:05:39.550  & +01:43:56.52  &   3.867  $\pm$ 0.008&  19.37  \\
SDSSJ1211+1224 & 12:11:46.935   & +12:24:19.08    &  3.862   $\pm$ 0.008  &19.97 \\  
SDSSJ1224+0746 &  12:24:20.658   &+07:46:56.33   &  3.867   $\pm$ 0.008& 19.08   \\
SDSSJ1258--0130 &  12:58:42.118   &-01:30:22.75   & 3.862     $\pm$ 0.008& 19.58   \\
SDSSJ2250--0846 &   22:50:52.659  & -08:46:00.22  & 3.869    $\pm$ 0.012 & 19.44   \\
SDSSJ2350+0025 &   23:50:32.306  & +00:25:17.23  &   3.876  $\pm$ 0.012&   20.61 
\enddata
\tablenotetext{}{\\}
\end{deluxetable*}

\subsection{Imaging on 17 Quasar Fields}
\label{ssec:obs}

Imaging observations for the sample of 17 quasar fields were carried out using the FOcal Reducer and low dispersion Spectrograph 2 (FORS2, \citealt{Appenzeller92}) instrument on the the Very Large Telescope (VLT) on 19 different nights between September, 2014 and March, 2015 (Program ID: 094.A-0900). The field-of-view of FORS2 is $6.8\times6.8$ arcmin$^{2}$ corresponding to $\sim 9.8\times9.8\,h^{-2}{\rm cMpc}^{2}$ at $z=3.88$. We used a $2\times 2$ binning readout mode, which results in an image pixel scale of $\rm 0.25 \arcsec/pix$. Each quasar field was observed using the NB HeI ($\lambda_{\rm c} = 5930$\AA, $\rm FWHM = 63$\AA) and the broad bands $g_{\rm HIGH}~(\lambda_{\rm c} = 4670$\AA, hereafter $g$) and $R_{\rm SPECIAL}~(\lambda_{\rm c} = 6550$\AA, hereafter $R$) in order to detect LAEs at $z\sim3.88$ (see top left panel of Fig.~\ref{fig:evoltracks}). 

The total exposure time per target for HeI, $R$, and $g$ was 3660s, 360s, and 900s respectively, which were broken up into
shorter individual exposures with dithering in order to fill the gap between the CCDs. Note that the full sequence of
exposures for a given object were not necessarily observed on the same night. The seeing for the 19 nights cover a range of $0.6 - 1.5\arcsec$ and the median seeing is $0.94\arcsec$.
      
\subsection{Data Reduction and Photometric Catalogs}
\label{ssec:redu_phot}

We performed the data reduction using standard IRAF\footnote{Image Reduction and Analysis Facility} tasks and our own custom codes
written in the Interactive Data Language (IDL). The reduction process included bias subtraction and flat fielding, which was performed based on twilight flats.

Given that individual science frames per field could have been observed in different nights, the photometric calibration was done before the final image stack was created. For the calibration of the $R$ images, Stetson photometric fields \citep{Stetson00} were observed in this band several times during the course of each night. We considered all the non-saturated photometric stars observed in the field and obtained their instrumental magnitude by using SExtractor \citep{Bertin96}. Then, we computed the zeropoint by comparing these magnitudes with the tabulated standard star $R$ magnitudes from the Stetson catalogs, after correcting them with the corresponding color term to take into account the difference between the Stetson and FORS2 $R$ filter curves\footnote{This correction in the Vega photometric system is given by,\\ $R = R_{\rm Stetson}-0.0095598(R_{\rm Stetson}-V_{\rm Stetson})$}. The final zeropoint was computed as the median of the zeropoint obtained from each individual star of the field. The final zeropoint varies slightly over different nights ($<1\%$), and the uncertainties in the zeropoints are typically $0.02$ mag. 

Spectrophotometric standard stars from different catalogs \citep{Oke90, Hamuy92, Hamuy94} were observed each night to calibrate the HeI and $g$ images. For each night, we computed the spectrophotometric star magnitude by convolving the HeI and $g$ filter transmission curve with the star spectra and then we compared this magnitude with the instrumental magnitudes obtained using SExtractor for the observed star in both HeI and $g$ bands. For 3/19 nights, no spectrophotometric standard star was observed in the $g$ band. For one of those nights, we had a Stetson photometric field observed in the $g$-band, which was then used to perform the flux calibration analogous to the $R$ image flux calibration\footnote{In this case the correction for the differences between the Stetson and FORS2 filter curves in the Vega photometric system is given by, $g = V_{\rm Stetson} + 0.630(B_{\rm Stetson} -V_{\rm Stetson}) - 0.124$}. For the other two nights, we performed the flux calibration by using the SDSS photometric catalogs of stars in the science fields. Specifically, we compared the SDSS $g$ magnitudes of several stars in the field with their instrumental magnitude measured on our science images, and then we computed the zeropoint for those nights.

\begin{figure*}
\centering{\epsfig{file=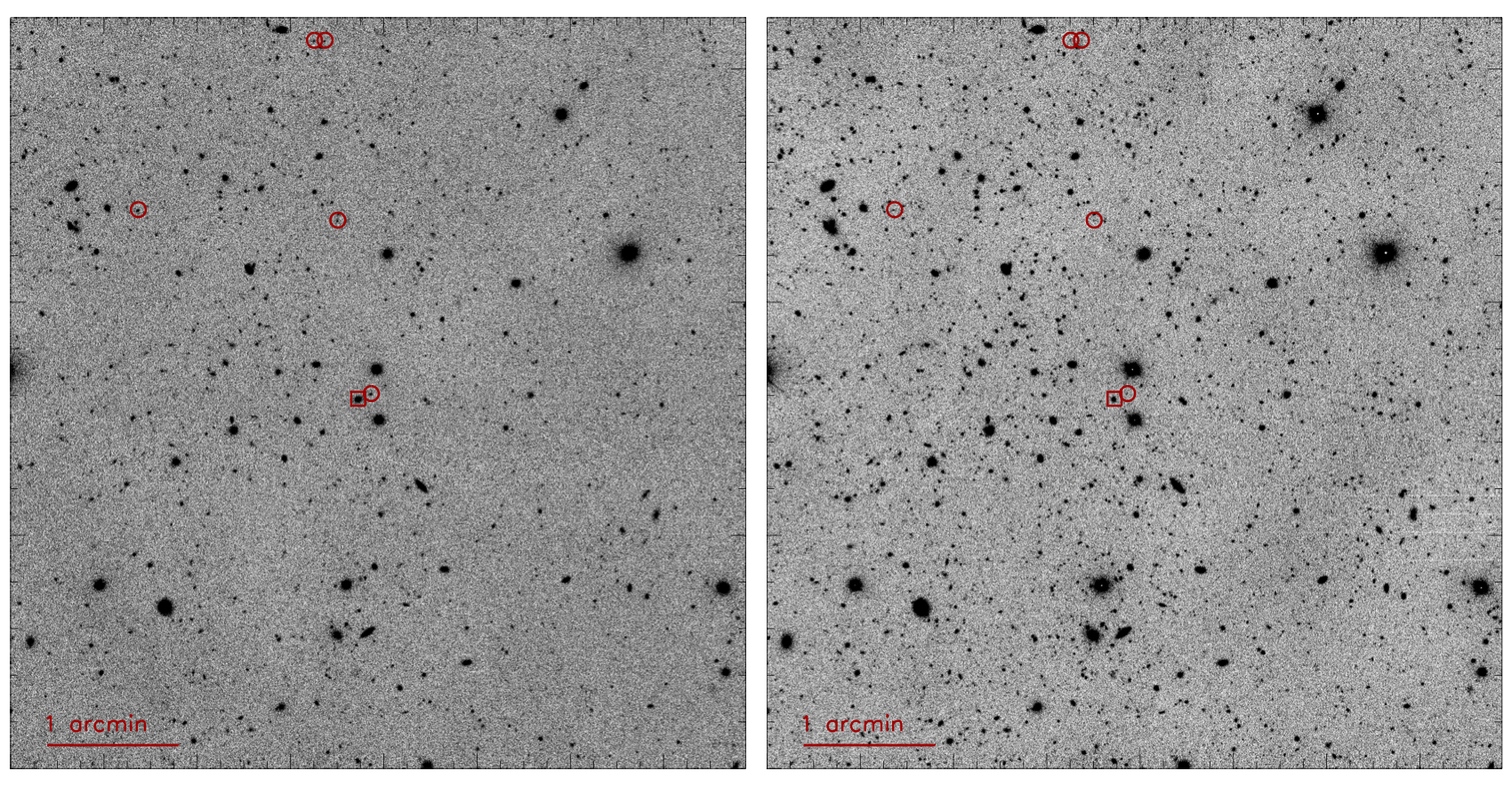, width=\textwidth}} 
\caption{\label{fig:0119image} HeI (left) and $R$ (right) reduced images of the SDSSJ0119--0342 field, centered on the quasar (red box). Detected LAEs are highlighted (red circles).\\}
\end{figure*}

We calibrated all the individual science exposures using the computed
zeropoints, and then we corrected them for extinction due to airmass
using the atmospheric extinction curve for Cerro Paranal
\citep{Patat11}, and for galactic extinction calculated using
\citet{Schlegel98} dust maps and the \citet{Cardelli89} extinction law
with R$_{V} = 3.1$. We used SCAMP \citep{Bertin06} to compute the
astrometric solution of each frame, using the SDSS-DR9 r-band catalogs
as the astrometric reference. The typical RMS obtained in the
astrometric calibration is $<0.15\arcsec$. Finally, the individual
images were sky-subtracted, re-sampled and median-combined using SWarp
\citep{Bertin02}. The noisy edges of the combined images were trimmed
out. Example reduced image in the HeI and $R$ filters is shown in Fig~\ref{fig:0119image}.

Object detection and photometry were performed using SExtractor in a
dual mode. Since we are interested in the detection of objects with a
strong Ly$\alpha$ emission line located in the core of our NB, we used
the HeI image for the object detection. In order to perform an
adequate measurement of the object colors, for each field we convolved
our images with a Gaussian kernel to match the point-spread function
(PSF) of the $R$, $g$ and HeI images such that their final PSF
corresponds to the PSF of the image with the worst seeing. In
Table~\ref{table:limit_mag} we indicate the seeing values for each field,
measured after the convolution.
We measured the object aperture
magnitudes using a fixed aperture of $2\arcsec$ diameter. Objects not
detected or detected with a signal-to-noise ratio (${\rm S\slash N}) <
2$ either in $g$ or $R$, were assigned the value of the corresponding
$2\sigma$ limiting magnitude, listed in Table~\ref{table:limit_mag}. Here, the ${\rm S\slash N}$ of each
object is computed as the ratio of the aperture flux measured by
SExtractor and the rms sky noise computed using our own IDL procedure,
which performs $2\arcsec$ aperture photometry in $\sim5000$ different
random positions in the image to compute a robust measurement of the
mean sky noise in each single aperture. The rms sky noise is then calculated as the standard
deviation of the distribution of the $\sim5000$  measured 
mean sky noise values. In
Table~\ref{table:limit_mag} we listed the $2\sigma$ and $5\sigma$
limiting magnitude measured for each field. The median of the $5\sigma$
limiting magnitude for our fields is 24.45, 25.14 and 25.81 for the HeI, $R$ and $g$ images respectively.

\begin{deluxetable*}{lllllllcc}
\tabletypesize{\scriptsize}
\tablecaption{$5\sigma$ and $2\sigma$ limiting magnitudes per field measured in a $2\arcsec$ diameter aperture. We also show the seeing measured on the HeI images and the seeing measured after matching the PSF of the images.\label{table:limit_mag}}
\tablewidth{0pt}
\tablewidth{0.8\textwidth}
\tablehead{
\colhead{Field}&
\colhead{HeI$_{5\sigma}$}&
\colhead{$R_{5\sigma}$}&
\colhead{$g_{5\sigma}$}&
\colhead{HeI$_{2\sigma}$}&
\colhead{$R_{2\sigma}$}&
\colhead{$g_{2\sigma}$}&
\colhead{seeing [$\arcsec$]}&
\colhead{seeing$_{\rm Final}$ [$\arcsec$]}
}
\startdata
SDSSJ0040+1706 & 24.35  &   25.14   &   25.72   &   25.35  &  26.13   &  26.72  & 0.86  & 1.24\\
SDSSJ0042--1020 & 24.40  &   25.08   &   25.79   &   25.39  &  26.08   &  26.78  & 1.12  & 1.26\\
SDSSJ0047+0423 & 24.50  &  25.15    &  25.83    &  25.50   & 26.15    & 26.82  & 1.22   & 1.37\\
SDSSJ0119--0342 & 24.37  &   25.11   &   25.86   &   25.37  &  26.10   &  26.85  & 0.68  & 0.75\\
SDSSJ0149--0552 & 24.45  &  25.11    &  25.92    &  25.45   & 26.11    & 26.92  & 0.72   & 0.97\\
SDSSJ0202--0650 & 24.40  &  25.16    &  25.85    &  25.39   & 26.15    & 26.85  & 1.17   & 1.17\\
SDSSJ0240+0357 & 24.60  &   25.14   &   25.70   &   25.59  &  26.13   &  26.70  & 0.98  & 0.98\\
SDSSJ0850+0629 & 24.48  &  25.17    &  25.76    &  25.48   & 26.17    & 26.75  & 0.83   & 1.00\\
SDSSJ1026+0329 & 24.54  &   24.92   &   25.85   &   25.54  &  25.91   &  26.85  & 1.08  & 1.11\\
SDSSJ1044+0950 & 24.51  &   25.15   &   25.93   &   25.50  &  26.14   &  26.92  & 0.90  & 1.21\\
SDSSJ1138+1303 & 24.40  &   24.84   &   25.66   &   25.39  &  25.83   &  26.66  & 0.86  & 1.07\\
SDSSJ1205+0143 & 24.52  &   25.19   &   25.91   &   25.51  &  26.19   &  26.91  & 1.13  & 1.49\\
SDSSJ1211+1224 & 24.37  &   24.78   &   25.70   &   25.36  &  25.78   &  26.70  & 0.65  & 0.71\\
SDSSJ1224+0746 & 24.43  &   24.91   &   25.67   &   25.42  &  25.90   &  26.66  & 1.02  & 1.16\\
SDSSJ1258--0130 & 24.43  &   25.15   &   25.88   &   25.42  &  26.15   &  26.87  & 0.96  & 1.24\\
SDSSJ2250--0846 & 24.55  &   25.06   &   25.81   &   25.54  &  26.06   &  26.81  & 0.87  & 1.02\\
SDSSJ2350+0025 & 24.45  &   25.17   &   25.72   &   25.44  &  26.17   &  26.71  & 0.61  & 0.63
\enddata
\tablenotetext{}{\\}
\end{deluxetable*}

\begin{figure}
\centering{\epsfig{file=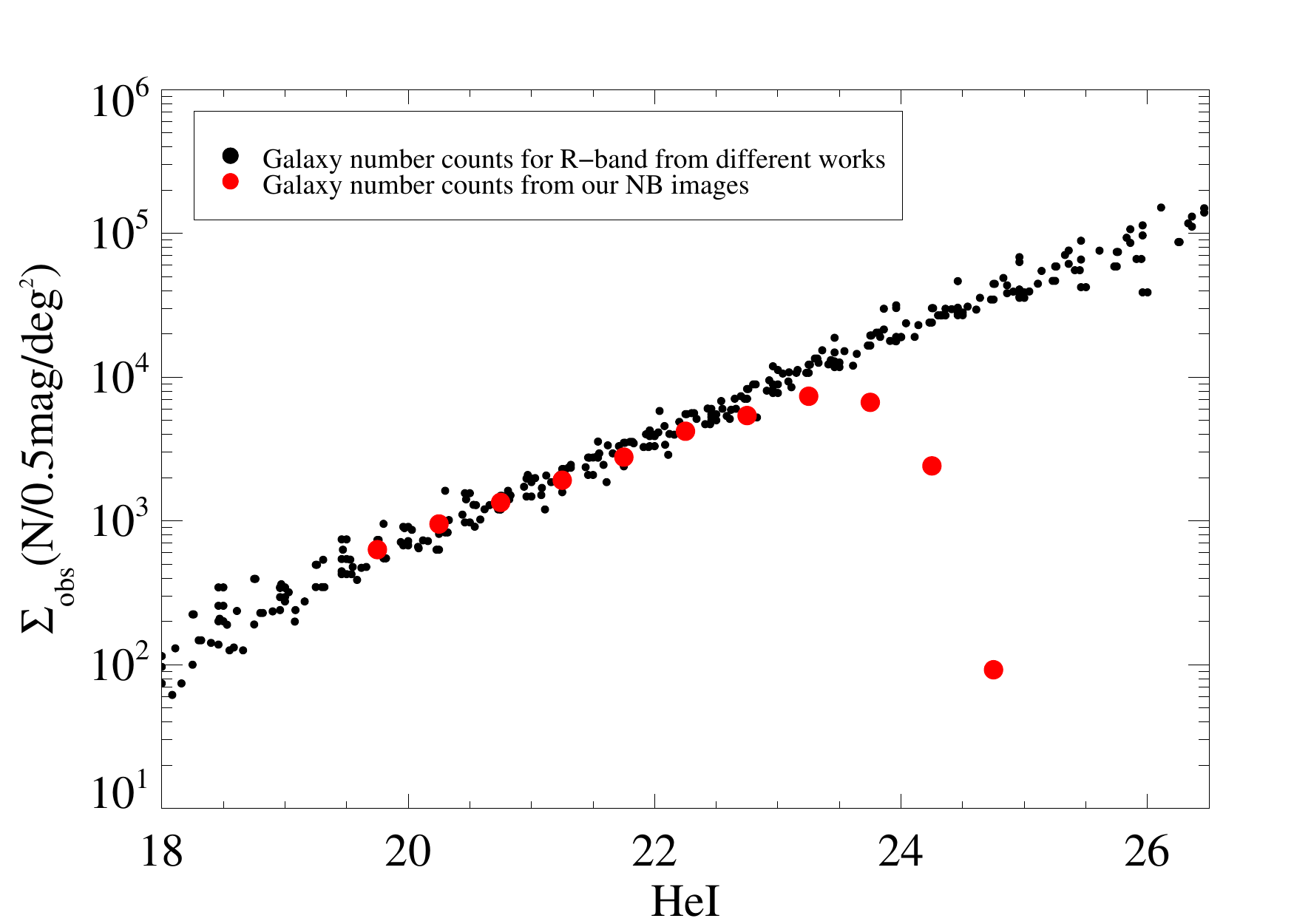, width=\columnwidth}} 
\caption{\label{fig:n_counts} HeI number counts detected in our NB images (red points) compared with the $R$-band number counts measured in previous studies (black points). In our NB images we find similar number counts as those detected in the $R$-band in previous studies, which validates the reduction process and flux calibration performed for the HeI filter.\\}
\end{figure}

One way to check if the reduction process and flux calibration is
correct for our NB filter is to compute the number counts from the HeI
images and compare it with the $R$-band number counts measured by 
previous work. The central wavelength of the HeI filter ($\lambda_{\rm c} = 5930$\AA) is sufficiently close to
the central wavelength of the $R$-band filter ($\lambda_{\rm c} = 6550$\AA), such that for sources whose
spectra do not vary strongly across the extend of the $R$-band filter,
the HeI magnitude is a good proxy for the $R$ magnitude.
Thus we 
expect that the HeI number counts should be a decent proxy for
the $R$-band number
counts measured previously. For this check, we used a
compilation of $R$-band number counts measured in 0.5 mag bins that includes number counts measured in the SDSS r-band as well as those
measured in the $R$ Kron-Cousins band
\citep{Metcalfe01,Shanks15}\footnote{The compilation of $R$-band number
  counts is available at
  \url{http://astro.dur.ac.uk/~nm/pubhtml/counts/counts.html}}. We compared
this with the number counts detected in all our fields in 0.5 mag bins
divided by the total effective area of our survey.  We show our
results in Fig.~\ref{fig:n_counts}, and we see that the HeI number
counts agree fairly well with the $R$-band number counts.  The
disagreement at the faintest magnitudes we probe is due to
incompleteness in the detection of sources in our images.

\begin{figure}
\centering{\epsfig{file=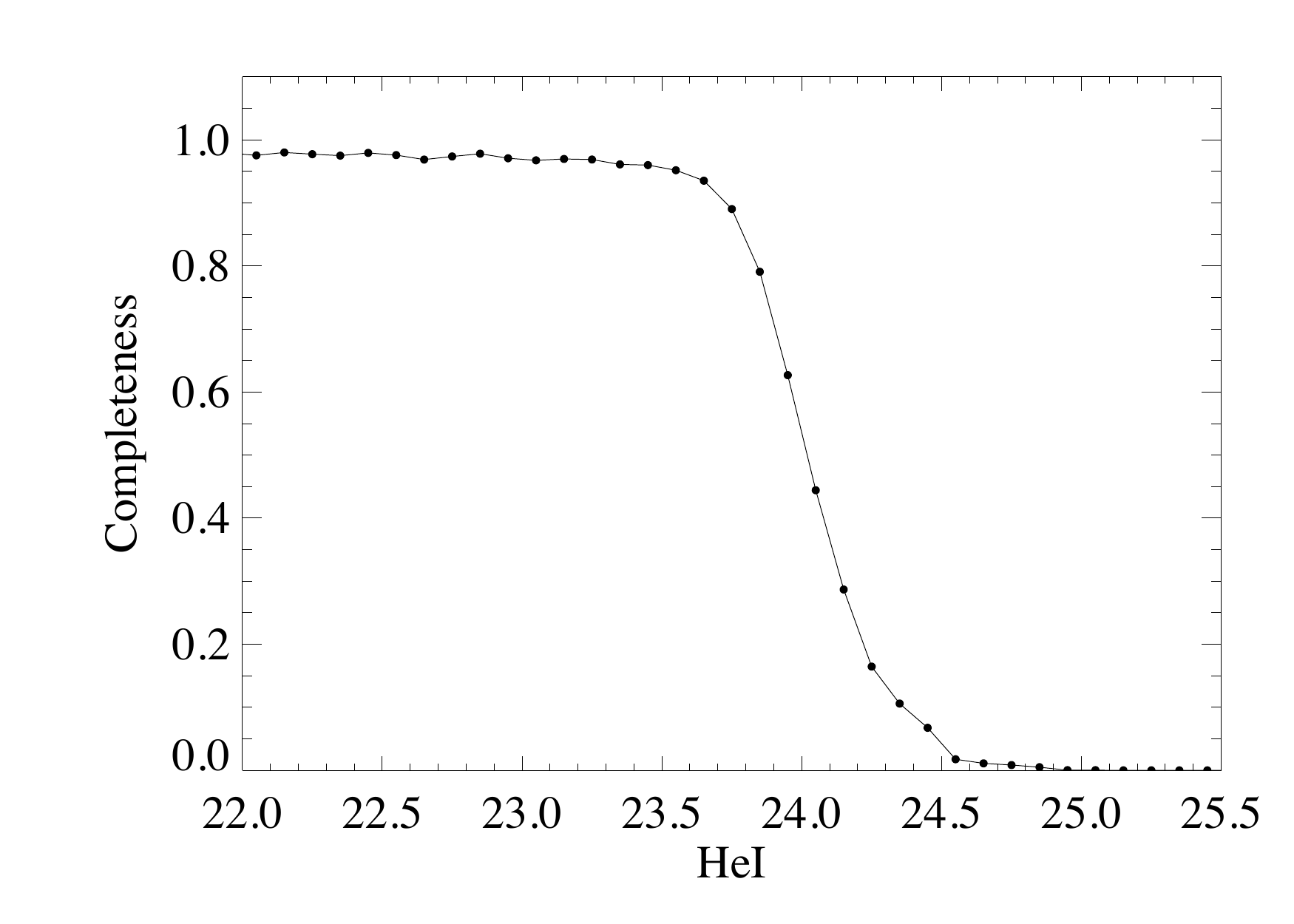, width=\columnwidth}} 
\caption{\label{fig:comple} Median of the completeness in the object detection on the HeI images as a function of the HeI magnitude. This completeness is computed for point sources.\\}
\end{figure}

We directly measured the completeness in the object detection in our
HeI images for each field as follows. Artificial point-sources with a
fixed total magnitude were inserted at random locations into the HeI images and then we ran SExtractor on these new images including artificial sources in the same way we did to create our photometric catalogs. The completeness per field was computed as the fraction of artificial sources that were actually detected. We repeated this procedure for different magnitudes bins spaced by 0.1. We computed the median of the completeness over our fields which is shown in Fig.~\ref{fig:comple}. Note that the object completeness depends on the size of the object, thus
our completeness computation is only valid for LAEs at $z\sim4$ which are assumed to be unresolved in our images (the typical effective radius of  $z\sim4$ LAEs is $r_{e} \sim 0.5$\, pkpc \citep[e.g.][]{Shibuya19} which corresponds to $0.07\arcsec$ at $z=4$).

\section{LAE Number Density in Quasar Fields}
\label{sec:number_density}

In this section we describe how LAEs were selected, we present the sample, and we compute the LAE number density in our fields to compare with the number counts in blank fields. 

\subsection{LAE Selection Criteria}
\label{ssec:selection}

As we discuss in the next section, clustering measurements are based on the comparison of the LAE number density detected in our fields to that expected at random locations in the universe. This last quantity should be preferentially computed either from blank fields observed using the same filter configuration as our quasar fields or from the outer parts of the images when the field-of-view is large enough such that the background level is reached. Given that we do not have control fields and that the field-of-view of FORS2 is not large enough for a determination of this background, the only alternative is to compute it from which the $z\sim4$ LAE luminosity function measured in previous studies. This may introduce systematic errors if our procedure for selecting LAEs differs from that used in previous work on blank fields from which the LAE luminosity
function has been estimated.  In order to limit the impact of such systematics and thus enable a robust clustering measurement, we must be careful to select
our LAEs following the same criteria adopted by previous studies. 

\begin{figure*}
\centering{\epsfig{file=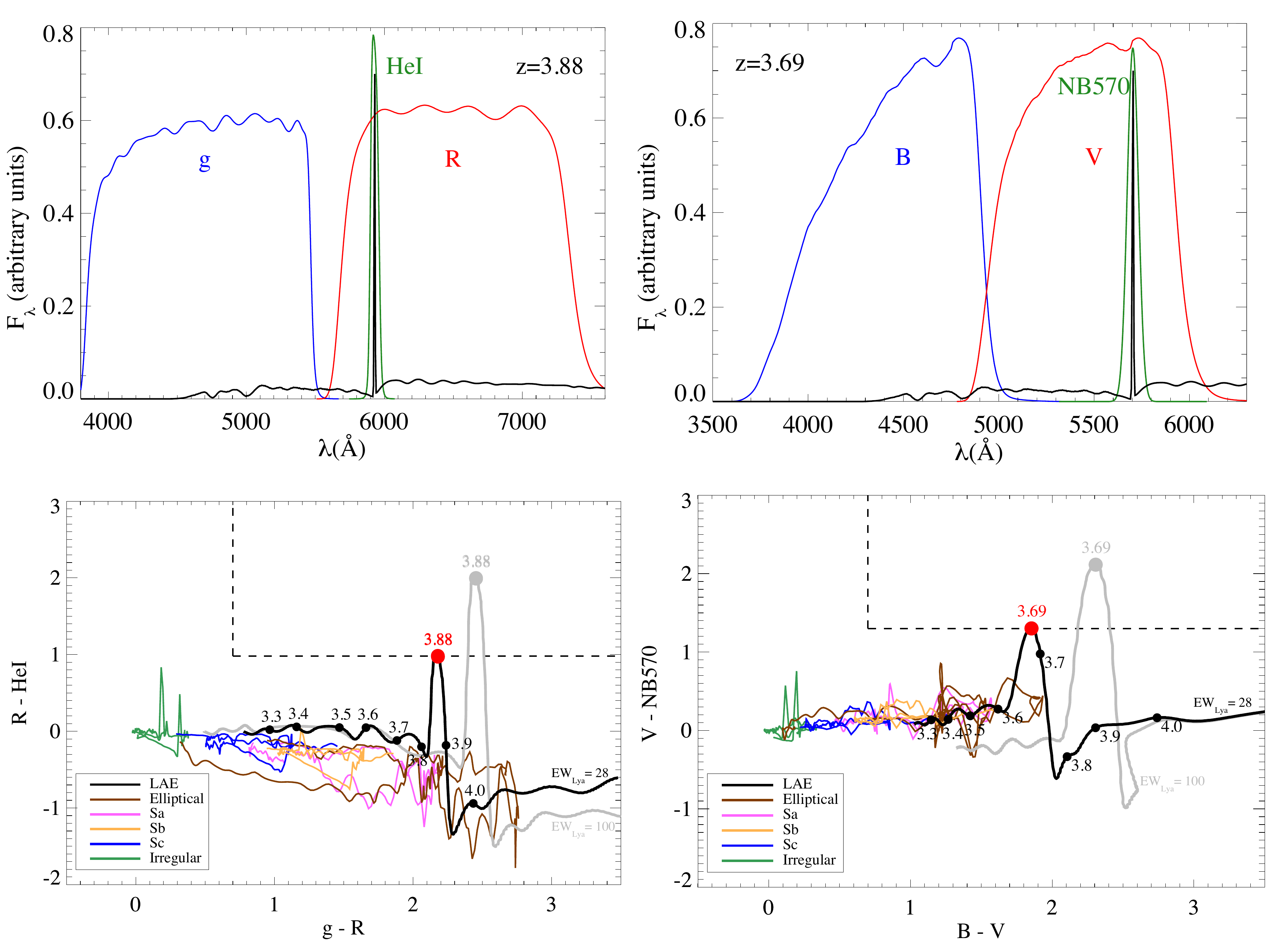, width=\textwidth}}
\caption{\textit{Top left:} Filter transmission curves used in this work shown on a simulated LAE spectra redshifted to $z\sim3.88$. The simulated spectra correspond to a galaxy with a flat UV continuum slope ($\propto\lambda^{-2}$) and with a $\rm Ly\alpha$ equivalent width of $28$\AA\, (in rest-frame). \textit{Bottom left:} Evolutionary track of a LAE with EW$_{\rm Ly\alpha}=28$\AA\, (black curve). Filled circles over the black curve indicate colors of LAEs from redshift 3.3 to 4.0, and the large red point indicate the exact position of the color of a LAE at $z=3.88$. For comparison, we also plotted the evolutionary track of a LAE with EW$_{\rm Ly\alpha}=100$\AA\, (gray curve). We overplotted the evolutionary tracks of other galaxies for our filter system redshifted from $z=0$ to $z=3$. We plot as brown, magenta, orange, blue, and red curves the evolutionary track of elliptical, Sa, Sb, Sc, and irregular galaxies respectively. The black dashed line show the selection region used in this work to select LAEs. \textit{Top right:} Filter transmission curves used in \citet{Ouchi08} shown on the simulated LAE spectra redshifted to $z\sim3.69$. \textit{Bottom right:} Same as the bottom left panel, but for the \citet{Ouchi08} filter system. In this case the large red point over the black curve indicate the exact position of the color of a LAE at $z=3.69$. The black dashed line show the selection region used by \citet{Ouchi08} to select LAEs.\\}
\label{fig:evoltracks} 
\end{figure*}

Here we use the LAE luminosity function measured by \citet{Ouchi08}, who selected $z=3.69$ LAEs using Subaru Suprime-Cam imaging data obtained in the filters $B$, $\rm NB570$ and $V$ analogously to our filter configuration $g$, HeI, and $R$ (see top right panel of Fig.~\ref{fig:evoltracks}). The LAE color selection criteria they used are defined by the equations $V-{\rm NB570}>1.3$ and $B-V>0.7$, thus we need to translate their color cuts to our $R - {\rm HeI}$ and $g - R$ colors respectively. If we perform our LAE selection exactly as they do, then we can ensure that the color completeness (i.e the completeness of the LAEs selection which does not have to be confused with the photometric completeness computed in ~\S~\ref{ssec:redu_phot}), contamination level, and the underlying background number density
of our LAE sample is the same as theirs.

The $V-{\rm NB570}$ color provides information about the observed-frame Ly$\alpha$ equivalent width (EW$_{\rm Ly\alpha}$) of the selected galaxy. Thus we must compute the limiting ${\rm EW}_{\rm Ly\alpha}$ value which
corresponds to the \citet{Ouchi08} color cut $V-{\rm NB570}>1.3$, and choose a color cut using our filters that isolates LAEs of the same
 ${\rm EW}_{\rm Ly\alpha}$. 
The observed-frame ${\rm EW}_{\rm Ly\alpha}$ is defined as:
\begin{equation}
{\rm EW}_{\rm Ly\alpha} = \int \frac{F_{\rm Ly\alpha}}{F_{\rm cont,Ly\alpha}} d\lambda, 
\label{eq:ew}
\end{equation}
where $F_{\rm Ly\alpha}$ is the specific flux of the ${\rm Ly\alpha}$ line
(with the continuum subtracted), and $F_{\rm cont,Ly\alpha}$ is the
specific flux of the continuum at the wavelength of the ${\rm
  Ly\alpha}$ line. The flux in the NB570 and the $V$ filters provides
information about the flux of the ${\rm Ly\alpha}$ line plus the
continuum covered by each filter curve, respectively. Note that the
shape of the continuum over the NB570 and $V$ bands is not trivial,
because there is a flux break at $\lambda_{\rm rest-frame}=1216$\AA\,
due to the absorption of photons with $\lambda_{\rm
  rest-frame}<1216$\AA\, by neutral hydrogen in the intergalactic
medium (see for example the LAE simulated spectra in the top-panels of Fig.~\ref{fig:evoltracks}). This makes it challenging to analytically compute both $F_{\rm
  Ly\alpha}$ and $F_{\rm cont,Ly\alpha}$ only from a $V-{\rm NB570}$
color information (the same complication applies to
using our $R$ and HeI filters to select LAEs at $z=3.88$). Our approach to determine the
EW$_{\rm Ly\alpha}$ from color measurements and determine
the equivalent set of color selection criteria in our filter is to
simulate LAE spectra with different EW$_{\rm Ly\alpha}$ values, and then
integrate them against the filters $V$ and NB570 to compute the colors
from the simulated galaxy spectra. Once we determine which set of model galaxy parameters
reproduce the $V-{\rm NB570}$ color value used
by \citet{Ouchi08}, we integrate the same spectra (now with known
EW$_{\rm Ly\alpha}$) against our filters $R$ and HeI, and we define
the $R- {\rm HeI}$ color criteria that we should use.

To simulate the LAE spectra, we start with a template star-forming
galaxy spectrum generated from a \citet{Bruzual03} stellar
population synthesis model\footnote{Obtained from \url{http://bruzual.org/}}, corresponding to
an instantaneous burst of star-formation with an age of 70Myr, a
\citet{Chabrier03} IMF, and a metallicity of 0.4Z$_{\odot}$. We
measured the UV slope of this template over the range 1300\AA
$<\lambda<$2000\AA\, by fitting a power law shape given by $A
\lambda^{\alpha}$, and then we modify its UV slope by multiplying by the
power law required to obtain a final flat UV continuum
($\propto\lambda^{-2}$), which is expected for LAEs at $z\sim4$
\citep{Overzier08, Ouchi08}. We then multiplied the resulting
spectrum at $\lambda\leq 912$\AA\, by an escape fraction parameter $f_{\rm
  esc}^{\lambda<912}=0.05$ to take into account the fact that only a
small fraction of Lyman limit photons are able to escape high-redshift
galaxies. To simulated the Ly$\alpha$ emission, we added
Gaussian line with rest-frame central wavelength $\lambda_{\rm Ly\alpha}=1215.7$\AA,  a $\rm FWHM_{\rm Ly\alpha}=1.05$\AA\, (or equivalently
$\rm 260\,km\,s^{-1}$) which agrees with measurements for
high-redshift LAEs \citep[e.g.][]{Venemans05, Shimasaku06}, and
amplitude $B$ which adjusts the intensity of the line in order to
model a Ly$\alpha$ line with a chosen rest-frame EW$_{\rm Ly\alpha}$ computed
using eqn.~(\ref{eq:ew})\footnote{Note that this simulated spectra is initially created at $z=0$, then the rest-frame and observed-frame EW$_{\rm Ly\alpha}$ are the same, and giving by eqn.~(\ref{eq:ew}).}. Finally, we redshifted the spectra to
$z=3.69$ (i.e., we are assuming that the line is in the center of the
NB filter) and attenuate the flux blueward of the ${\rm Ly\alpha}$
line using an IGM transmission model $T_{z}(\lambda)$ from
\citet{Worseck11}, which models the foreground Lyman series absorption
from the IGM. Note that we only attenuate the continuum flux, but we do not attenuate the flux of the ${\rm Ly\alpha}$
line. In this way, the chosen rest-frame EW$_{\rm Ly\alpha}$ value in our simulation correspond to an apparent EW$_{\rm Ly\alpha}$, which is assumed to be already affected by the IGM absorption, and then this is comparable to the EW$_{\rm Ly\alpha}$ values measured directly from observed LAEs. This is different from the intrinsic EW$_{\rm Ly\alpha}$ which is the EW$_{\rm Ly\alpha}$ after corrected for the effect of foreground IGM absorption. We simulate spectra with different rest-frame EW$_{\rm
  Ly\alpha}$ values and integrated them against the filters used by
\citet{Ouchi08} to compute $V-{\rm NB570}$. We find that a LAE with
rest-frame EW$_{\rm Ly\alpha}=28$\AA\, reproduces the color
criteria used by \citet{Ouchi08} of  $V - {\rm NB570}=1.3$\footnote{Note that \citet{Ouchi08} compute an apparent EW$_{\rm Ly\alpha}\sim44$\AA\, for their LAE selection, which differs from the value determined from our modeling at the 36\% level. This difference could be attributed to the different UV continuum slope used in their modeling,  and differences in the IGM transmission models adopted (they use models from \citet{Madau95}).}. We
then redshifted such spectra up to $z=3.88$ (using the corresponding
redshift dependent IGM transmission model) and integrated it against our filters, finding
$R - {\rm HeI}=0.98$, which defines the color criteria to select LAEs
using our filter configuration.

We show an example of a simulated spectra at $z=3.88$ in the top left panel of Fig.~\ref{fig:evoltracks}. We also used our simulated spectra to compute an evolutionary track of LAEs in the color-color diagram. To this end, we redshifted it from $z=3.2$ to $z=4.4$ with a spacing of 0.02, applied the corresponding IGM transmission model in each redshift step, and integrated the spectrum against both our and \citet{Ouchi08} filter curves. The evolutionary tracks are shown in the bottom left panel of Fig.~\ref{fig:evoltracks} for the $g$, HeI, $R$ filter set and in the bottom right panel of Fig.~\ref{fig:evoltracks} for the $B$, $\rm NB570$ and $V$ filter set.

The second color criteria used by \citet{Ouchi08} to select LAEs is $B-V>0.7$, which guarantees a low level of low-redshift galaxy contamination in
the resulting sample (they estimate a contamination level of
$0-14\%$).  We studied the $B-V$ colors of typical LAEs by using our
simulated spectra with EW$_{\rm Ly\alpha}=28$\AA. We modified the UV
continuum slope of the spectra in order to span the range of observed
values in $z=4$ galaxies which goes from $\alpha=-3$ to $\alpha=0$
\citep{Bouwens09}, and we found that in this range the $B-V$ color
varies from $1.7$ to $2.2$. Given our parametrization of the escape fraction and the
IGM absorption, a simulated LAE would have a color of $B-V=0.7$
only if their spectra had an UV continuum slope of $\alpha=-8.8$, which is
unphysically steep. 
Therefore, typical LAEs always satisfy
the $B-V$ condition, and it is chosen mostly to exclude low-redshift
galaxy contamination.

We study the colors of low-redshift galaxies in both the $B$, $V$ and
the $g$, $R$ filter system in order to compare how different they
are. We used a set of five galaxy templates, typically used to
estimate photometric redshifts. The templates are from the photo-$z$
code EASY \citep{Brammer08}, which are distilled from the PEGASE
spectral synthesis models. We redshifted these templates from $z=0$ to
$z=3$, and integrated them against the $B$, NB570, $V$, and the $g$,
HeI, $R$ filter transmission curves to generate their evolutionary
track in the color-color diagram. We show our results in
the bottom panels of Fig.~\ref{fig:evoltracks}. Given that low-redshift galaxies have similar $B-V$ and
$g-R$ colors, we simply adopt the same color cut as in \citet{Ouchi08} for
our filters, given by $g-R>0.7$. The locus of the low-redshift
galaxies is well isolated from the location of the LAEs we wish to select,
and the $g-R=0.7$ color cut helps exclude strong OIII emitters at
$z=0.18$ or OII emitters at $z=0.59$, represented by the two peaks in
the green curves in the bottom panels of Fig.~\ref{fig:evoltracks}.

To summarize,
the color cuts used in this work are thus:
\begin{eqnarray}
& R - {\rm HeI} & > 0.98  \nonumber \\
&  g - R & > 0.7 
\label{eq:color_eq}
\end{eqnarray}

\subsection{LAE Sample}
\label{ssec:sample}

\begin{figure*}
\centering{\epsfig{file=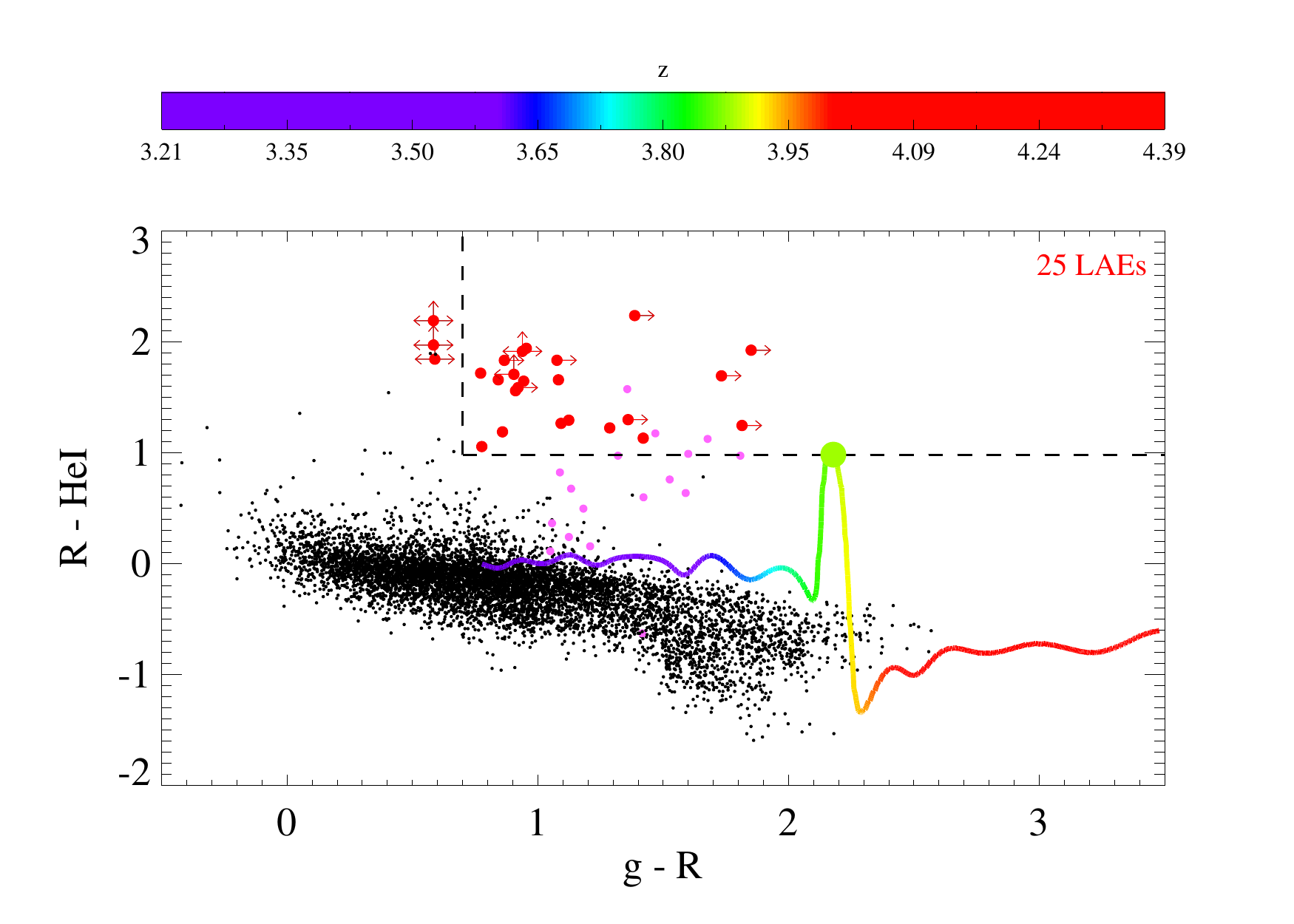, width=0.8\textwidth}}
\caption{Color-color diagram for our 17 stacked quasar fields. Here the LAE evolutionary track showed in the bottom left panel of Fig.~\ref{fig:evoltracks} is plotted as redshift color-coded track according to the color bar. Magenta points indicate the color of the quasar in our filters. Arrows indicate lower limits for the colors, and then double arrows represent cases in which the object was not detected at 2$\sigma$ level in either $g$ and $R$ images, and each magnitude was replaced by the corresponding limit magnitude. The dashed line is indicating the selection region defined by eqns.~(\ref{eq:color_eq}). We have highlighted the selected LAEs of our main sample as red points.\\}
\label{fig:colorcolor_together}
\end{figure*}

We used our photometric catalogs to select LAE candidates in our 17
quasar fields. In order to ensure a solid detection in the NB filter, we
only selected objects detected with ${\rm S\slash N}\ge 5.0$ in the
HeI band. We required that sources have the SExtractor parameter $\rm
FLAGS=0$ in order to exclude blended, saturated or truncated
sources. Additionally, we created masks from our images to indicate
``bad regions", which are defined as regions where the object
detection is not reliable. Some examples of such regions are
locations in the vicinity of extremely bright stars  (since contamination from
bright objects affect the photometry of nearby sources), highly extended
objects which cover a large area of the image (which may preclude the
detection of background LAEs), and satellite trails. 
We discarded all the sources located on the defined bad
regions. Note that these masks are also used in the clustering
analysis presented in section ~\S~\ref{sec:clustering} to compute the
effective area of our survey. All the objects in our 17 quasar fields
satisfying the aforementioned requirements are plotted in the color-color
diagram shown in Fig.~\ref{fig:colorcolor_together}.

\begin{figure*}
\centering{\epsfig{file=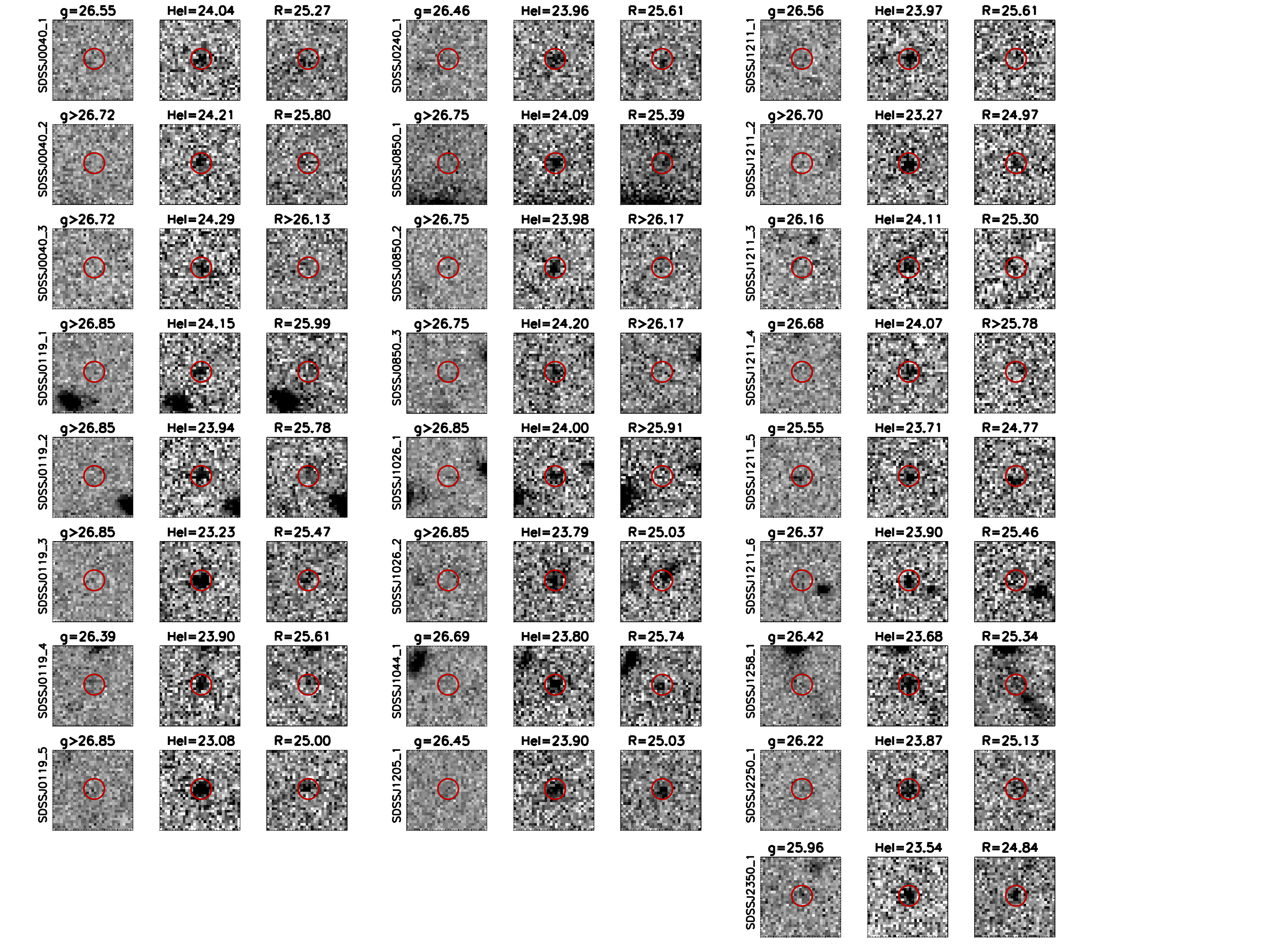, width=1.15\textwidth}}
\caption{$g$, HeI, and $R$ images of all the LAE sample, exhibited in panels of $7.5\arcsec\times7.5\arcsec$. A red circle of $2\arcsec$ in diameter shows the position of the detected LAEs. Magnitudes are indicated in each panel. \\}
\label{fig:LAE_images}
\end{figure*}

\begin{figure}
\centering{\epsfig{file=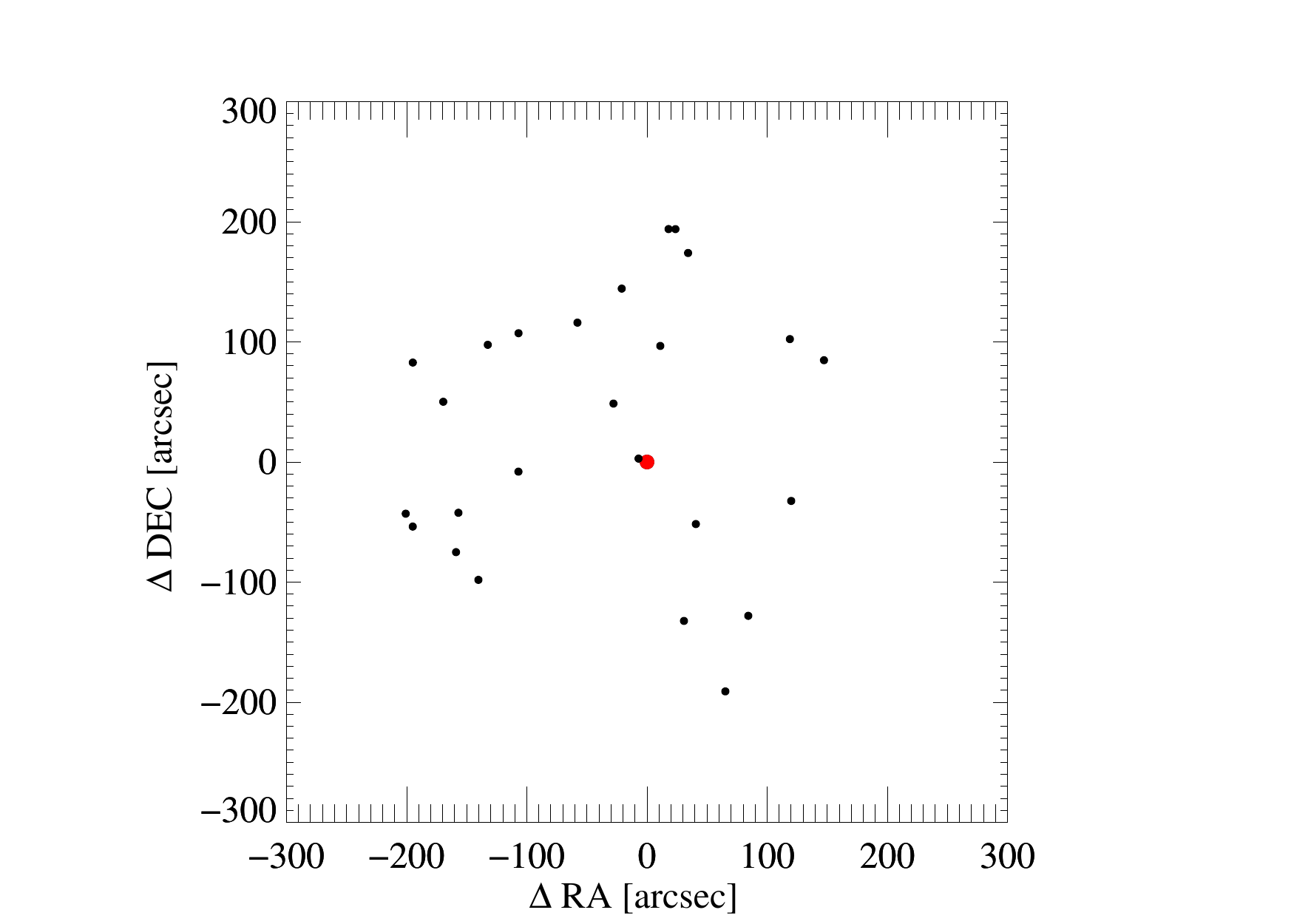, width=\columnwidth}}
\caption{Distribution of the 25 LAEs with respect to the central quasar plotted as a red dot. \\}
\label{fig:LAE_dist}
\end{figure}

All the objects satisfying the color criteria in
eqns.~(\ref{eq:color_eq}) were selected as LAEs (this selection region
is shown as a dashed line in Fig.~\ref{fig:colorcolor_together}). 
We discarded objects with $\rm HeI<22.7$ in order to exclude bright
low-redshift interloper galaxies that might be affecting our selection. This lower limit is the same used by \citet{Ouchi08}. 
Note that data points with double arrows in the
x-axis of this figure represent cases in which the object was not
detected at the 2$\sigma$ level in either $g$ and $R$ images, and each
magnitude was replaced by the corresponding limiting magnitude in each
field. Thus the position on the g-R axis of these points is determined by
the  depth of our $g$ and $R$ images, but does not provide
any actual information  about the $g-R$ color. We note that the
broad band limiting magnitude of the \citet{Ouchi08} work
are deeper than ours (their $V$ images are $\sim$2.0 mag
deeper than our corresponding depth in $R$, and $B$ images
are $\sim$1.8 mag deeper than our $g$ depth). If we had deeper $R$ images left pointing arrows would move
to the left and upward pointing arrows would move up in Fig.~\ref{fig:colorcolor_together}, then we could lose some of the selected LAEs, but we would not gain objects. On the other hand, deeper $g$ images would move right pointing arrows to the right, and we could only gain objects. Specifically there are three additional objects that could satisfy the color criteria (the objects with double arrows in the $g$-$R$ direction outside of our selection region in Fig.~\ref{fig:colorcolor_together}), and for this reason, we also considered them as LAEs. We recall that the $g-R>0.7$ color cut is chosen mostly to
mitigate low-redshift interlopers, and for such
interlopers we expect a solid detection in both $g$ and $R$ at the 2$\sigma$ limiting
magnitude of our observations. The fact that these three additional objects are not detected in our broad bands suggests that are more likely to be LAEs. Our final sample, comprises 25 LAEs whose photometry is shown
in Table~\ref{table:LAEsample}. We show cutout images of the detected LAEs in
Fig.~\ref{fig:LAE_images}. We also show the distribution of LAEs
around the central quasar for all the fields
together in  Fig.~\ref{fig:LAE_dist}.

\begin{deluxetable*}{l r r r r r}
\tabletypesize{\scriptsize}
\tablecaption{Properties of LAE candidates. The magnitudes correspond to AB magnitudes measured in a $2\arcsec$ diameter aperture for each band. \label{table:LAEsample}\\}
\tablewidth{0pt}
\tablewidth{0.8\textwidth}
\tablehead{
\colhead{ID}&
\colhead{RA}&
\colhead{DEC}&
\colhead{$R$}&
\colhead{$g$}&
\colhead{HeI}\\
\colhead{}&
\colhead{(J2000)}&
\colhead{(J2000)}&
\colhead{}&
\colhead{}&
\colhead{}
}
\startdata
SDSSJ0040$+$1706\_1& 10.04149& 17.10328& 25.27& 26.55&24.04\\
SDSSJ0040$+$1706\_2& 10.03181& 17.07825& 25.80&$>$26.72&24.21\\
SDSSJ0040$+$1706\_3& 10.01593& 17.09058&$>$26.13&$>$26.72&24.29\\
SDSSJ0119$-$0342\_1& 20.00311& -3.65073& 25.99&$>$26.85&24.15\\
SDSSJ0119$-$0342\_2& 20.00470& -3.65074& 25.78&$>$26.85&23.94\\
SDSSJ0119$-$0342\_3& 19.99616& -3.70381& 25.47&$>$26.85&23.23\\
SDSSJ0119$-$0342\_4& 20.00119& -3.67776& 25.61& 26.39&23.90\\
SDSSJ0119$-$0342\_5& 20.03121& -3.67618& 25.00&$>$26.85&23.08\\
SDSSJ0240$+$0357\_1& 40.09660&  3.92959& 25.61& 26.46&23.96\\
SDSSJ0850$+$0629\_1&132.57429&  6.44336& 25.39&$>$26.75&24.09\\
SDSSJ0850$+$0629\_2&132.57962&  6.46084&$>$26.17&$>$26.75&23.98\\
SDSSJ0850$+$0629\_3&132.52617&  6.52619&$>$26.17&$>$26.75&24.20\\
SDSSJ1026$+$0329\_1&156.67078&  3.48840&$>$25.91&$>$26.85&24.00\\
SDSSJ1026$+$0329\_2&156.58314&  3.52040& 25.03&$>$26.85&23.79\\
SDSSJ1044$+$0950\_1&161.07159&  9.83493& 25.74& 26.69&23.80\\
SDSSJ1205$+$0143\_1&181.42337&  1.69562& 25.03& 26.45&23.90\\
SDSSJ1211$+$1224\_1&182.93760& 12.41883& 25.61& 26.56&23.97\\
SDSSJ1211$+$1224\_2&182.89731& 12.41925& 24.97&$>$26.70&23.27\\
SDSSJ1211$+$1224\_3&182.88843& 12.39337& 25.30& 26.16&24.11\\
SDSSJ1211$+$1224\_4&182.95711& 12.39097&$>$25.78& 26.68&24.07\\
SDSSJ1211$+$1224\_5&182.92907& 12.43754& 24.77& 25.55&23.71\\
SDSSJ1211$+$1224\_6&182.90785& 12.43242& 25.46& 26.37&23.90\\
SDSSJ1258$-$0130\_1&194.71644& -1.48276& 25.34& 26.42&23.68\\
SDSSJ2250$-$0846\_1&342.72901& -8.71839& 25.13& 26.22&23.87\\
SDSSJ2350$+$0025\_1&357.62878&  0.46156& 24.84& 25.96&23.54
\enddata
\tablenotetext{}{}
\end{deluxetable*}

\subsection{Comparison with the Number Density of LAEs in Blank Fields}
\label{ssec:compar_bkg}

Here we compute the total number density of LAEs in our quasar fields and compare it with the LAE number density measured in blank fields. Given that we matched the LAE selection criteria with that used in \citet{Ouchi08}, we directly compare our number density with theirs. 

\begin{figure}
\centering{\epsfig{file=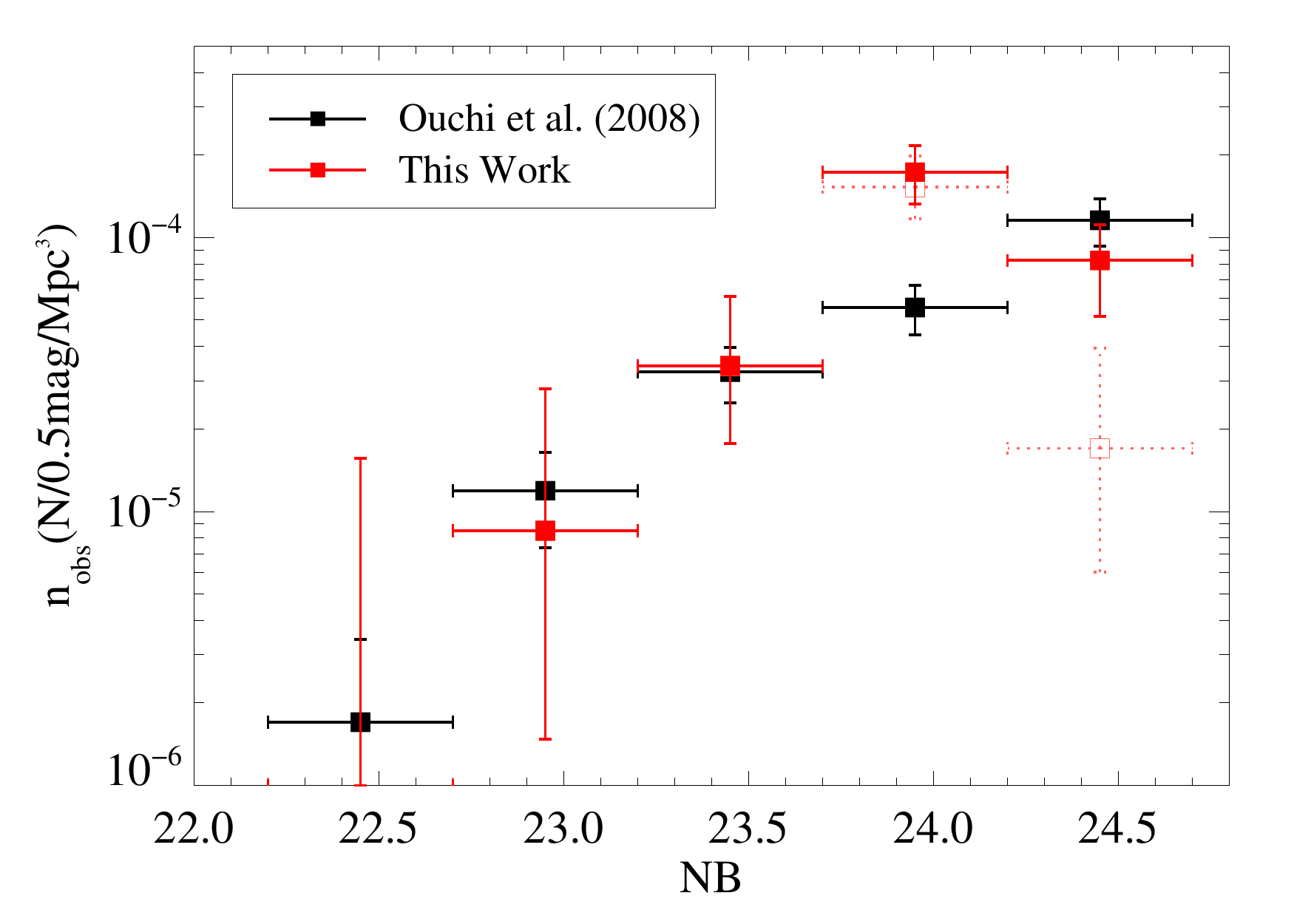, width=\columnwidth}}
\caption{LAEs number density for our 17 stacked fields. The black squares are the LAEs number density per NB570 magnitude range in blank fields \citep{Ouchi08}, the filled red squares are the completeness-corrected LAEs number density per HeI magnitude range in quasar fields computed using our LAE sample, with error bars computed using one-sided Poisson confidence intervals for small number statistics from \citet{Gehrels86}. Open red squares are this measurement before to consider the completeness detection. We detect an overdensity of LAEs in quasar fields compared with that measured in blank fields.\\}
\label{fig:n_lae}
\end{figure}

We compute the number density of LAEs, by dividing the number of observed LAEs per HeI magnitude bin by the effective survey volume. For the computation of the effective survey volume, we assumed a top-hat function for our NB filter curve transmission function, with width equal to the FWHM (63\AA), and computed the comoving distance coverage at $z=3.88$, resulting 37.4 cMpc. We multiply this quantity by the effective area of our survey, which is computed per field by subtracting the
masked area from the total area of the image. The total area of our
survey is obtained by adding the effective area of the 17 fields
(listed in Table \ref{table:field_param}). We obtain that our survey
covered a total area of 740.8 arcmin$^{2}$ corresponding to 3,146 cMpc$^{2}$. We compared our measurement with the results
from \citet{Ouchi08} in Fig.~\ref{fig:n_lae} (we have converted the surface number counts presented in \citet{Ouchi08} to the volume number counts by taking into account the FWHM of their NB filter transmission curve (69\AA), which corresponds to a comoving distance coverage of 43.2 cMpc at $z=3.7$).

Note that all our quasar fields reach different NB limiting magnitude (with a median $5\sigma$ limiting magnitude of $24.45$) and have different object completeness detection.  We have also computed the completeness-corrected number density by using the completeness we determined
per field as described in \S~\ref{ssec:redu_phot} which is shown in Fig.~\ref{fig:n_lae}. We recall that the HeI magnitude is not
providing information about the Ly$\alpha$ line flux, but rather
about the sum of the Ly$\alpha$ line flux and the
continuum flux. We find that the number density of LAEs in quasar fields
is consistent with that measured in blank fields for the brightest NB magnitudes bins but we detected a significantly higher number density
(by a factor of 3.1) for LAEs with $23.7<NB<24.2$, the range in which most of our objects (18 out of 25) are. We find agreement with the
\citet{Ouchi08} number density for the faintest magnitude bin, which only contains 2 of our LAEs, but we note that our sample is highly
incomplete at those magnitudes (completeness of $\sim$10\%, as shown in Fig.~\ref{fig:comple}), thus our completeness-corrected number density could be dominated by errors in the completeness computation in this magnitude range. Overall, our results suggest that we have  detected an overdensity of LAEs in quasar environments on the scales probed by this work ($R\lesssim7\,h^{-1}\,{\rm cMpc}$).

\section{Clustering Measurements}
\label{sec:clustering}

In the last section we showed that quasar fields have a higher number
density of LAEs than blank field pointings. By measuring the
cross-correlation and auto-correlation function of LAEs in our survey,
we can further quantify the clustering of LAEs around quasars and
determine their spatial profile.  To measure the clustering of LAEs in
quasar fields, we perform the same analysis presented in
\citet{garciavergara17}. Here we only describe the most important
points and refer the reader to that paper for further details.

\subsection{Quasar-LAE Cross-correlation Function}
\label{ssec:cross}
We
measure a volume-averaged projected cross-correlation function between
quasars and LAEs defined by
\begin{equation}
\chi (R_{\rm min}, R_{\rm max}) = \frac{\int \xi_{QG} (R,Z)  dV_{\rm eff}}{V_{\rm eff}}, 
\label{eq:estimator2_simpler}
\end{equation}
where $\xi_{QG} (R,Z)$ is the real space quasar-LAE cross-correlation function and $V_{\rm eff}$ is the effective volume of the survey, defined as a cylindrical volume with transverse separation from $R_{\rm min}$ to $R_{\rm max}$ and height $Z$ which is the radial comoving width probed by our survey. We measured the volume-averaged projected cross-correlation function in logarithmically spaced radial bins centered on the quasar by using the estimator,
\begin{equation}
\chi (R_{\rm min}, R_{\rm max})=  \frac{\langle QG\rangle}{\langle QR\rangle} -1 
\label{eq:estimator}
\end{equation}
where $\langle QG\rangle$ is the number of quasar-LAE pairs in the bin, which is directly measured by counting the quasar-LAE pairs found in our images, and $\langle QR\rangle$ is the expected number of quasar-LAE pairs in the same bin if they were randomly distributed around the quasar, with the background number density.
The quantity  $\langle QR\rangle$ is computed from:
\begin{equation}
\langle QR\rangle =  n_{\rm G}(z,{L_{\rm Ly\alpha}>L_{\rm Ly\alpha}^{\rm limit}}) V_{\rm eff}, 
\label{eq:NLBG2}
\end{equation}
where $n_{\rm G}(z,{L_{\rm Ly\alpha}>L_{\rm Ly\alpha}^{\rm limit}})$ is the mean number density of LAEs at redshift $z$ with $\rm Ly\alpha$ luminosity greater than the limiting $\rm Ly\alpha$ luminosity reached in our survey. Here, we use the LAE luminosity function measured by \citet{Ouchi08} at $z=3.7$,
  which is given by the Schechter parameters $\phi^{*}=3.4\times10^{-4}$\,Mpc$^{-3}$, $L^{*}_{\rm Ly\alpha}=1.2\times10^{43}$\,erg\,s$^{-1}$ mag, and $\alpha = -1.5$. We compute $n_{\rm G}(z,{L_{\rm Ly\alpha}>L_{\rm Ly\alpha}^{\rm limit}})$ for each individual quasar field by integrating the luminosity function up to the 5$\sigma$ limiting $\rm Ly\alpha$ luminosity of
the field in question. Note that converting from the 5$\sigma$
limiting HeI magnitude (presented in section \S~\ref{ssec:redu_phot})
to the 5$\sigma$ limiting $\rm Ly\alpha$ luminosity is not trivial
because the HeI magnitude includes both the flux of the $\rm Ly\alpha$
line and the flux of the continuum. For this conversion, we have used
our simulated LAE spectra (created as we describe in section
\S~\ref{ssec:selection}) to compute the $\rm Ly\alpha$ limiting
luminosity corresponding to a given HeI limiting
magnitude. Specifically, we simulate a flat UV continuum LAE spectrum
($\propto\lambda^{-2}$) with both a fixed $\rm EW_{Ly\alpha}=28$\AA\,
and without the $\rm Ly\alpha$ emission line. Then, we determine the
flux that both spectra would have in the HeI filter, and then subtract
them to obtain the $\rm Ly\alpha$ luminosity. In other words, the
resulting $\rm Ly\alpha$ luminosity depends on the assumed $\rm
EW_{Ly\alpha}$ value, and the properties of the simulated spectrum. We
choose the $\rm EW_{Ly\alpha}$ value here to match the limiting
equivalent width of our LAE selection. 
Considering our limit $\rm
EW_{Ly\alpha}=28$\AA\, the $\rm Ly\alpha$ luminosity could be
sightly underestimated if the candidates typically have greater $\rm
EW_{Ly\alpha}$ values (specifically, if we consider LAEs with $EW_{Ly\alpha}=80$\AA\, the $\rm Ly\alpha$ luminosity would change in a $\sim8\%$), however, the errors associated with the Schechter
parameters of the LAE luminosity function dominate over this
relatively small source of error in the
LAE number density computation. The mean 5$\sigma$ limiting $\rm
Ly\alpha$ luminosity computed for our fields is listed in Table
\ref{table:field_param}, on average we obtained $L_{\rm Ly\alpha}=4.1\times10^{42}$\,erg\,s$^{-1}$. Given that the source
detection in our images is not 100\% complete up to the 5$\sigma$
limiting $\rm Ly\alpha$ luminosity (see section
\S~\ref{ssec:redu_phot}), we have included the completeness correction in the
computation of $n_{\rm G}(z,{L_{\rm Ly\alpha}>L_{\rm Ly\alpha}^{\rm limit}})$
by weighting the luminosity function by the source detection
completeness computed as explained in section \S~\ref{ssec:redu_phot}
for each field. We listed the $n_{\rm G}(z,{L_{\rm Ly\alpha}>L_{\rm Ly\alpha}^{\rm limit}})$ values per field in Table
\ref{table:field_param}. 

For the computation of the $V_{\rm eff}$ value in eqn.~(\ref{eq:NLBG2}), we considered the volume of the bin defined by a cylinder with radial width $(R_{\rm max} - R_{\rm min})$ and height $Z$ which is computed from the FWHM of our NB filter (we approximate the filter curve transmission function as a top-hat function with a width equal to the FWHM). Specifically, for HeI, the $\rm FWHM = 63$\AA\, corresponds to a redshift coverage of $\Delta z=0.052$
at $z=3.88$ (or equivalently $\rm 3,197\,km\,s^{-1}$)  and a comoving distance of
$Z =26.2\,h^{-1}$\,cMpc. We also considered an angular selection function in this computation, which is estimated using the detection masks created from our images (see section \S~\ref{ssec:sample}). These masks quantify the fraction of the bin area where the LAEs were detectable. In Table \ref{table:field_param} we show the effective volume of each field $V_{\rm field}$ (i.e the sum of the $V_{\rm eff}$ over the radial bins). We obtained that the total volume of our survey is 40,254 $\,h^{-3}$\,cMpc$^{3}$.

\begin{deluxetable*}{l r c r r r c c}
\tabletypesize{\small}
\tabletypesize{\scriptsize}
\tablecaption{LAE number counts in each individual quasar field.\label{table:field_param}}
\tablewidth{0pt}
\tablewidth{0.8\textwidth}
\tablehead{
\colhead{Field}&
\colhead{$A_{\rm field}$}&
\colhead{${\rm log}(L_{\rm Ly\alpha}^{\rm limit})$}&
\colhead{$n_{\rm G}$}&
\colhead{$V_{\rm field}$}&
\colhead{$\langle QR\rangle_{\rm field}$}&
\colhead{$\langle QG\rangle_{\rm field}$}&
\colhead{$\delta$}\\
\colhead{(1)}&
\colhead{(2)}&
\colhead{(3)}&
\colhead{(4)}&
\colhead{(5)}&
\colhead{(6)}&
\colhead{(7)}&
\colhead{(8)}
}
\startdata	
SDSSJ0040$+$1706 & 42.56 & 42.65 &  0.49 & 2310.95 &   1.13 &  3 &  2.66\\
SDSSJ0042$-$1020 & 44.31 & 42.64 &  0.25 & 2407.74 &   0.60 &  0 &  0.00\\
SDSSJ0047$+$0423 & 45.05 & 42.59 &  0.25 & 2446.96 &   0.60 &  0 &  0.00\\
SDSSJ0119$-$0342 & 44.72 & 42.64 &  0.60 & 2430.48 &   1.46 &  5 &  3.41\\
SDSSJ0149$-$0552 & 44.42 & 42.61 &  0.64 & 2415.15 &   1.54 &  0 &  0.00\\
SDSSJ0202$-$0650 & 41.44 & 42.64 &  0.21 & 2255.96 &   0.47 &  0 &  0.00\\
SDSSJ0240$+$0357 & 44.48 & 42.56 &  0.45 & 2416.79 &   1.08 &  1 &  0.92\\
SDSSJ0850$+$0629 & 41.06 & 42.60 &  0.56 & 2235.92 &   1.25 &  3 &  2.41\\
SDSSJ1026$+$0329 & 45.04 & 42.58 &  0.38 & 2444.89 &   0.93 &  2 &  2.16\\
SDSSJ1044$+$0950 & 45.01 & 42.59 &  0.52 & 2444.62 &   1.28 &  1 &  0.78\\
SDSSJ1138$+$1303 & 37.68 & 42.64 &  0.45 & 2047.16 &   0.92 &  0 &  0.00\\
SDSSJ1205$+$0143 & 43.37 & 42.59 &  0.32 & 2356.25 &   0.75 &  1 &  1.34\\
SDSSJ1211$+$1224 & 45.05 & 42.64 &  0.65 & 2446.81 &   1.60 &  6 &  3.75\\
SDSSJ1224$+$0746 & 44.34 & 42.62 &  0.35 & 2410.69 &   0.84 &  0 &  0.00\\
SDSSJ1258$-$0130 & 44.12 & 42.62 &  0.39 & 2396.02 &   0.93 &  1 &  1.08\\
SDSSJ2250$-$0846 & 44.82 & 42.58 &  0.53 & 2434.21 &   1.30 &  1 &  0.77\\
SDSSJ2350$+$0025 & 43.32 & 42.61 &  0.72 & 2353.88 &   1.69 &  1 &  0.59\\
All &  740.81 &  &  &  40254.46 &  18.36 & 25 &  1.36
\enddata
\tablecomments{
(1) Field ID, 
(2) Total area of the field in units of arcmin$^{2}$,
(3) 5$\sigma$ limiting $\rm Ly\alpha$ luminosity in units of log$($\,erg\,s$^{-1})$, 
(4) The mean number density of $z\sim4$ LAEs in units of $(10^{-3}\,h^{3}$\,cMpc$^{-3})$, for $L_{\rm Ly\alpha}>L_{\rm Ly\alpha}^{\rm limit}$. Given that $L_{\rm Ly\alpha}^{\rm limit}$ and the source detection completeness are different for each field, we obtain a number density slightly different for each one, 
(5) Total volume of the field in units of $(h^{-3}$\,cMpc$^{3})$, 
(6) Total number of expected LAEs on the whole field. This is computed as $\langle QR\rangle_{\rm field} = n_{\rm G}V_{\rm field}$,
(7) Total number of observed LAEs on the whole field,
(8) Total overdensity per field, computed as $\langle QG\rangle_{\rm field}/\langle QR\rangle_{\rm field}$\\ 
}
\end{deluxetable*}

In Table \ref{table:field_param} we also list the values of $\langle
QG\rangle_{\rm field}$ and $\langle QR\rangle_{\rm field}$ which
correspond to the sum of the $\langle QG\rangle$ and $\langle
QR\rangle$ values respectively over the bins for each individual quasar
field. This provides a measurement of the individual overdensity
$\delta=\langle QG\rangle _{\rm field}/ \langle QR\rangle_{\rm
  field}$. We find that 7 out of 17 fields have more LAEs than the
expected number in blank fields (i.e $\delta>1.0$). If we sum over
all the $\langle QR\rangle _{\rm field}$ values for our 17 fields we obtain
that one would expect to detect a total of $\sim 18.36$ LAEs for blank field
pointings, whereas we detected a total of 25 LAEs, meaning that on average
$z\sim4$ quasars fields are overdense in LAEs by a factor of $\sim
1.4$. 

We emphasize the fact that our conclusions are based on the average number density of 17 quasar fields. We note that 10 of our fields are indeed underdense, but they are not representative of the aggregate behavior of the quasars in our sample. Those 10 fields serve as a reminder of the large cosmic variance inherent
in quasar environment studies, and the danger of over-interpreting noisy measurements made from studies where only a single quasar or a handful of fields are analyzed
at the highest ($z\gtrsim6$) redshifts. 

\begin{figure}
\centering{\epsfig{file=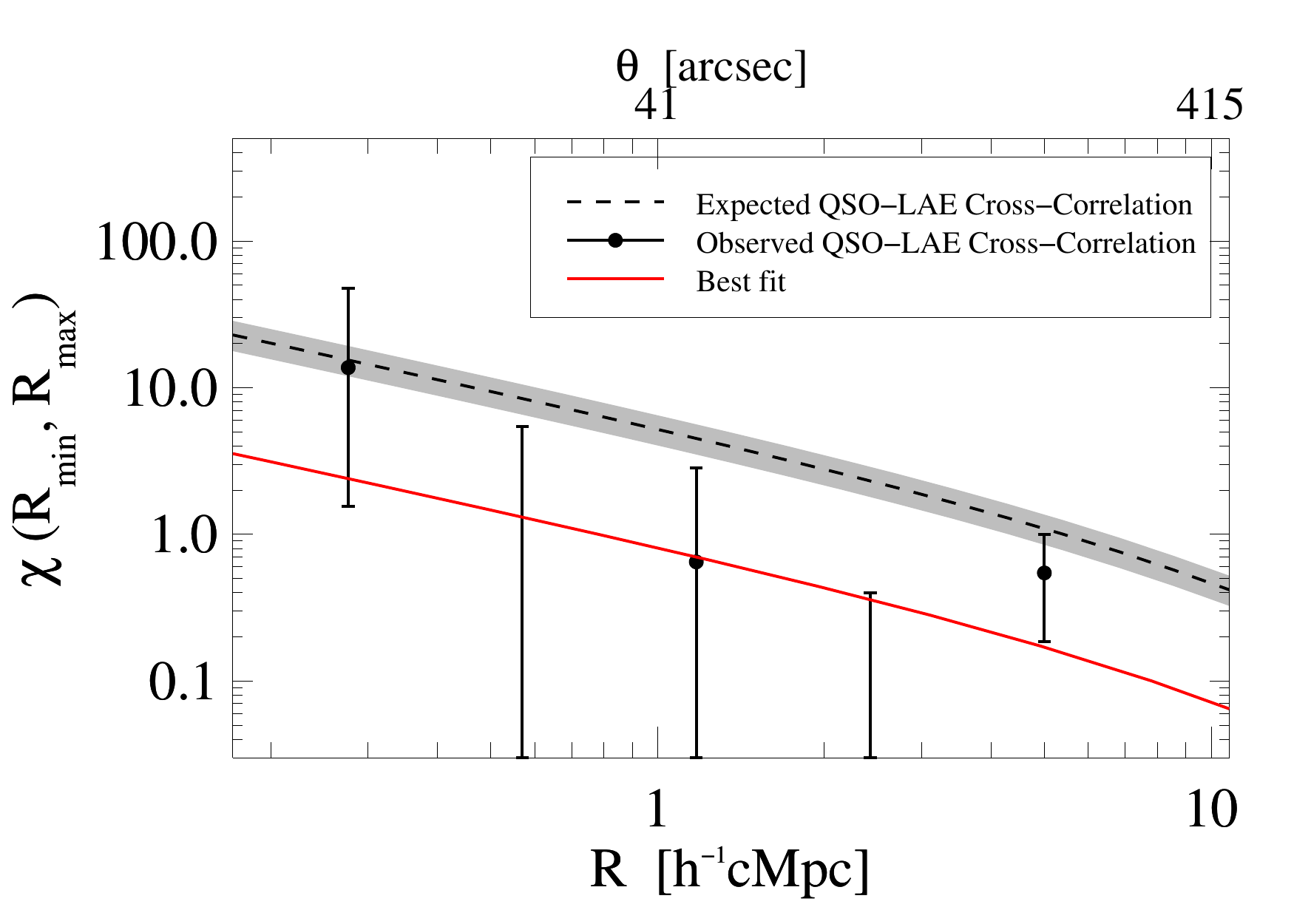, width=\columnwidth}}
\caption{Quasar-LAE cross-correlation function. The filled circles are showing our measurement with $1\sigma$ Poisson error bars and the red curve shows the best fit for our measurement given by $r^{QG}_{0}=2.78^{+1.16}_{-1.05} \,h^{-1}\,{\rm cMpc}$ for a fixed slope at $\gamma=1.8$. The dashed black line shows the theoretical expectation of $\chi$ for our 17 stacked fields computed from the quasar and LAEs auto-correlation functions and assuming a deterministic bias model. The gray shaded region indicates the $1\sigma$ error on the theoretical expectation computed by error propagation and based on the reported $1\sigma$ errors of the $r^{QQ}_{0}$ and $r^{GG}_{0}$ parameters.\\}
\label{fig:crosscorr_LAE}
\end{figure}

We measure the quasar-LAE cross-correlation function by adding up the
$\langle QG\rangle$ and $\langle QR\rangle$ values, computed in bins of
transverse distance, for the 17 quasar fields and plugging those values
into  eqn.~(\ref{eq:estimator}). We show our results in
Fig.~\ref{fig:crosscorr_LAE} and tabulate the values in Table
\ref{table:chi_LAE}. Given the small size of our LAE sample, we have
assumed that Poisson error dominates our measurement, and in
Fig.~\ref{fig:crosscorr_LAE} we have plotted the one-sided Poisson
confidence interval for small number statistics from
\citet{Gehrels86}. Notwithstanding the large errors,
we do detect a positive quasar-LAE cross-correlation function which is consistent with a power-law shape indicative of a concentration of
LAEs centered on the quasars. 

\begin{deluxetable*}{c c r r r r}
\tabletypesize{\small}
\tabletypesize{\scriptsize}
\tablecaption{quasar-LAE Cross-Correlation Function.\label{table:chi_LAE}} 
\tablewidth{0pt}
\tablewidth{0.8\textwidth}
\tablehead{
\colhead{$R_{\rm min}$}&
\colhead{$R_{\rm max}$}&
\colhead{$\langle QG\rangle$}&
\colhead{$\langle QR\rangle$}&
\colhead{$\chi (R_{\rm min}, R_{\rm max})$}&
\colhead{$V_{\rm eff, total}$}\\ 
\colhead{$(h^{-1}$\,cMpc)}&
\colhead{$(h^{-1}$\,cMpc)}&
\colhead{}&
\colhead{}&
\colhead{}&
\colhead{$(h^{-3}$\,cMpc$^{3})$}
}
\startdata
  0.178& 0.368&           1& 0.068&13.719$^{+33.854}_{-12.173}$ & 149.06 \\
 0.368& 0.759&           0& 0.287&-1.000$^{+ 6.408}_{- 0.000}$  & 628.82\\
 0.759& 1.566&           2& 1.213& 0.649$^{+ 2.175}_{- 1.065}$  &2656.29\\
 1.566& 3.231&           4& 5.126&-0.220$^{+ 0.617}_{- 0.373}$  & 11237.91\\
 3.231& 6.664&          18&11.645& 0.546$^{+ 0.457}_{- 0.361}$  &25538.90
  \enddata
\tablecomments{Volume-averaged projected cross-correlation function between quasars and LAEs $\chi (R_{\rm min}, R_{\rm max})$ defined according to eqn.~(\ref{eq:estimator}) and measured in radial bins defined by $R_{\rm min}$ and $R_{\rm max}$ for our 17 stacked fields. We show the observed number of quasar-LAE pairs per bin $\langle QG\rangle$ and the expected number of quasar-random pairs per bin $\langle QR\rangle$ computed according to eqn.~(\ref{eq:NLBG2}). The total volume of the bin added over the fields is also listed.\\}
\end{deluxetable*}

In order to determine the real-space cross-correlation parameters
$r_{0}^{QG}$ and $\gamma$ that best fit our data, we use a maximum
likelihood estimator assuming a Poisson distribution for the number
counts and fit following the same procedure described in
\citet{garciavergara17}. Given the large errors in our measurements,
we fix the slope at $\gamma=1.8$ and find that the maximum
likelihood and the $1\sigma$ confidence interval for the correlation
length is $r^{QG}_{0}=2.78^{+1.16}_{-1.05} \,h^{-1}\,{\rm cMpc}$. We
used this value in eqn.~(\ref{eq:estimator2_simpler}) to compute the
corresponding $\chi (R_{\rm min}, R_{\rm max})$ value which is shown
as the red line in Fig.~\ref{fig:crosscorr_LAE}. In order to verify
that our results are not sensitive to our chosen binning, we have
also 
measured the cross-correlation function by changing the binning and
again fitting, and we find that different binning choices only change the constrained $r^{QG}_{0}$ parameter within the uncertainties quoted above.

For comparison we have computed the expected value for the quasar-LAE
cross-correlation function assuming a deterministic bias model. In
other words, if $\delta_{G}$ and $\delta_{Q}$ are the density contrast
of galaxies and quasars respectively, the cross-correlation between
quasars and LAEs is defined by $\xi_{QG} = \langle \delta_{Q}
\delta_{G} \rangle$.
Assuming that LAEs and quasars trace the same
underlying dark matter overdensities in a deterministic way,
the galaxy and quasar density contrast can be
written as $\delta_{G} = b_{G}(\delta_{DM})\delta_{DM}$ and
$\delta_{Q}= b_{Q}(\delta_{DM})\delta_{DM}$ respectively, with
$b_{G}(\delta_{DM})$ and $b_{Q}(\delta_{DM})$ the galaxy and quasar
bias respectively, which are (possibly non-linear) functions of the dark matter density
contrast, $\delta_{DM}$.
If we think of $\delta_{Q}$ and $\delta_{G}$ as two
stochastic processes, their cross-correlation coefficient can be
written as 
\begin{equation}
\rho= \frac{\langle  \delta_{Q} \delta_{G} \rangle}{\sqrt{\langle \delta_{Q}  \delta_{Q} \rangle \langle \delta_{G} \delta_{G}\rangle}}.  
\label{eq:cc_coef}
\end{equation}
But given the deterministic relations for $\delta_G$ and $\delta_Q$ above
it must be the case that $\rho$ in eqn.~(\ref{eq:cc_coef}) is equal one. In then
trivially follows that the quasar-LAE cross-correlation function can
be written  $\xi_{QG}=\sqrt{\xi_{QQ} \xi_{GG}}$, where $\xi_{QQ}$ and $\xi_{GG}$
are the auto-correlation of quasars and LAEs respectively. If we also
assume that $\xi_{QQ}$ and $\xi_{GG}$ have a power law shape given by
$\xi = (r/r_0)^\gamma$, with the same slope $\gamma$ for quasars and
LAEs, then the quasar-LAE cross-correlation length can be written as
$r^{QG}_{0} = \sqrt{ r^{QQ}_{0} r^{GG}_{0}}$. We used 
values for the auto-correlation lengths for LAEs and quasars reported at $z\sim4$, which are given
by $r^{QQ}_{0} = 22.3 \pm 2.5\,h^{-1}$\,cMpc for a fixed $\gamma=1.8$
\citep{Shen07}\footnote{\citet{Shen07} fitted the quasar auto-correlation function using a fixed $\gamma=2.0$ and reported an auto-correlation length of $r^{QQ}_{0} = 24.3 \pm 2.4\,h^{-1}$\,cMpc. The auto-correlation length $r^{QQ}_{0} = 22.3 \pm 2.5\,h^{-1}$\,cMpc used in our work, was obtained by fitting the \citet{Shen07} quasar auto-correlation measurement using a fixed $\gamma=1.8$.} and $r^{GG}_{0} = 2.74^{+0.58}_{-0.72}\,h^{-1}$\,cMpc also with
fixed $\gamma=1.8$ \citep{Ouchi10}.
This yields an expected cross-correlation length of
$r^{QG}_{0}=7.82 \pm 1.03\,h^{-1}$\,cMpc ($\gamma=1.8$), which is 2.8 higher than the cross-correlation length measured in our fields. Indeed, this model predicts that we should have detected 52.5 LAEs in the total volume of our survey which is
$2.1$ times higher than the 25 LAEs we actually detected. We plot the expected cross-correlation function as a dashed line in Fig.~\ref{fig:crosscorr_LAE}, which is 6.4 times higher than our fitted measurement (the red line in Fig.~\ref{fig:crosscorr_LAE}).
The deterministic bias model, which assumes only that quasars and LAEs probe
the same underlying dark matter overdensities without any additional
sources of stochasticity, appears to be  inconsistent with our measurements. 

The fact that an overestimation by a factor of 2.1 in the number of galaxies yields to an overestimation by a factor of 6.4 in the cross-correlation can be easily understood from our equations. If we consider the eqn.~(\ref{eq:estimator}), with $R_{\rm min}$ and $R_{\rm max}$ being the minimum $R$ value of the smaller bin in our measurements and the maximum $R$ value of the larger bin respectively, then $\langle QG\rangle$ and $\langle QR\rangle$ will be the total number of galaxies around the quasar and  expected in blank fields respectively, over all the physical scales traced in our work  (then we call them $\langle QG\rangle_{\rm total}$ and $\langle QR\rangle_{\rm total}$). The observed ratio between $\langle QG\rangle_{\rm total}$ and $\langle QR\rangle_{\rm total}$ is simply the overdensity we detected, which is $1.4$ (see table~\ref{table:field_param}), then the observed correlation function is $\chi_{\rm obs} = 0.4$. On the other hand, using the eqn.~(\ref{eq:estimator}) we can write the ratio between the expected to the observed correlation function as
\begin{equation}
\frac{\chi_{\rm exp}+1}{\chi_{\rm obs} +1} =  \frac{\langle QG\rangle_{\rm total, exp}}{\langle QG\rangle_{\rm total, obs}} 
\label{eq:chi_QG}
\end{equation}
where the subscripts $\rm exp$ and $\rm obs$ have been used to refer to the expected and observed quantities. The terms $\langle QR\rangle_{\rm total}$ have been cancel out because in our computation, $\langle QR\rangle_{\rm total}$ is not the observed number of galaxies in blank fields, but the expected one. Given that $\chi_{\rm obs}=0.4$, and considering this is underestimated by a factor of 6.4, the expected correlation function is $\chi_{\rm exp}=6.4\times0.4=2.56$, and the term $(\chi_{\rm exp}+1) / (\chi_{\rm obs} +1)$ in eqn.~(\ref{eq:chi_QG}) would be 2.5. Therefore, the right hand of eqn.~(\ref{eq:chi_QG}), which represent the overestimation in the number of galaxies would be given by this value. The value 2.5 is sightly higher than the overprediction of 2.1 that we computed, but this is because we computed the overprediction on the cross-correlation (by a factor of 6.4) based on the ratio between the expected and the fitted correlation function (instead of using the data points), which yields to small differences.

\subsection{LAE Auto-correlation Function in Quasar Fields}
\begin{figure}[t!]
\begin{center}
\centering{\epsfig{file=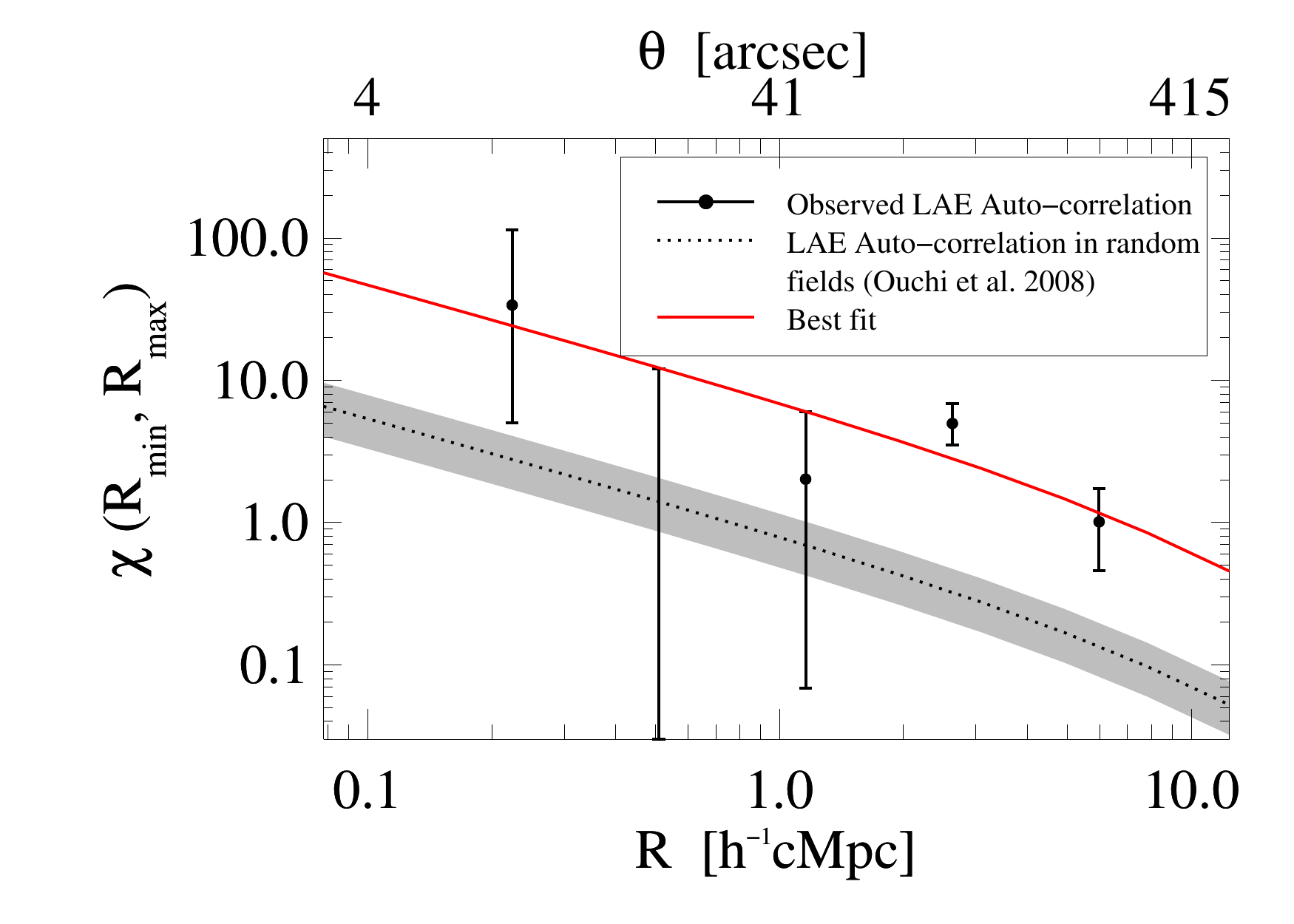, width=\columnwidth}}
\caption{LAE auto-correlation function in quasar fields. The filled circles are showing our measurement with $1\sigma$ Poisson error bars and the red curve shows the best fit for our measurement given by $r^{GG}_{0} = 9.12^{+1.32}_{-1.31}\,h^{-1}$\,cMpc for a fixed slope at $\gamma =1.8$. The dotted black line shows the observed LAE auto−correlation in blank fields at $z\sim4$ measured by \citet{Ouchi10} with the gray shaded area indicating the 1$\sigma$ errors based on their $1\sigma$ error reported for $r^{GG}_{0}$. We find that LAEs are significantly more clustered in quasar fields compared with their clustering in blank fields.\\}
\label{fig:autocorr_LAE}
\end{center}
\end{figure}

We also measured the LAE auto-correlation function in our fields, in order to compare it with that measured in blank fields. If quasars reside in overdensities of LAEs, then we expect the LAE auto-correlation to be significantly enhanced compared with blank fields \citep[see e.g.][]{garciavergara17}.
We measure it by using the estimator 
\begin{equation}
\chi(R_{\rm min}, R_{\rm max}) = \frac{\langle GG\rangle}{\langle RR\rangle} -1
\label{eq:estimator_auto}
\end{equation}
where $\langle GG\rangle$ is the observed number pairs of LAEs,
computed counting LAE pairs per radial bin in our images, and $\langle
RR\rangle$ is the number of LAE pairs that we expect to detect in
blank fields, assuming that galaxies were randomly distributed. This
last quantity has been computed as in eqn.~(13) of
\citet{garciavergara17}. Our results are shown in
Fig.~\ref{fig:autocorr_LAE}, and tabulated in Table
\ref{table:auto}. As for the cross-correlation measurement, error bars
on $\chi (R_{\rm min}, R_{\rm max})$ are computed using the one-sided
Poisson confidence intervals for small number statistics. 

\begin{deluxetable*}{c c r r r}
\tabletypesize{\small}
\tabletypesize{\scriptsize}
\tablecaption{LAEs Auto-Correlation Function.\label{table:auto}}
\tablewidth{0pt}
\tablewidth{0.8\textwidth}
\tablehead{
\colhead{$R_{\rm min}$}&
\colhead{$R_{\rm max}$}&
\colhead{$\langle GG\rangle$}&
\colhead{$\langle RR\rangle$}&
\colhead{$\chi (R_{\rm min}, R_{\rm max})$}\\
\colhead{$(h^{-1}$\,cMpc)}&
\colhead{$(h^{-1}$\,cMpc)}&
\colhead{}&
\colhead{}&
\colhead{}
}
\startdata
    0.136&  0.308&           1&  0.029& 33.760$^{+ 79.947}_{- 28.746}$  \\
  0.308&  0.700&           0&  0.142& -1.000$^{+ 12.980}_{-  0.000}$  \\
  0.700&  1.591&           2&  0.663&  2.018$^{+  3.981}_{-  1.950}$  \\
  1.591&  3.614&          16&  2.680&  4.970$^{+  1.895}_{-  1.477}$  \\
  3.614&  8.211&          13&  6.460&  1.012$^{+  0.728}_{-  0.551}$  
\enddata    
\tablecomments{LAE auto-correlation function in quasars fields $\chi (R_{\rm min}, R_{\rm max})$ shown in Fig.~\ref{fig:autocorr_LAE}.\\}
\end{deluxetable*}

Analogous to what we performed for the cross-correlation, we used
a maximum likelihood estimator for a Poisson distribution to fit our auto-correlation function. We fix the slope at $\gamma=1.8$ and find a correlation length of $r^{GG}_{0} = 9.12^{+1.32}_{-1.31}\,h^{-1}$\,cMpc plotted as a red line in Fig.~\ref{fig:autocorr_LAE}.  

The measured value for the volume-averaged auto-correlation function
at $z\sim 4$ is compared to our measurements in
Fig.~\ref{fig:autocorr_LAE}.  Here we plugged in the best-fit LAE
auto-correlation function parameters measured at $z\sim4$ by
\citet{Ouchi10} ($r^{GG}_{0}=2.74^{+0.58}_{-0.72} \,h^{-1}\,{\rm cMpc}$ and
$\gamma=1.8$) into eqn.~(\ref{eq:estimator2_simpler}) using a power
law form for $\xi_{GG} (R,Z)$, which gives the dotted line plotted in
Fig.~\ref{fig:autocorr_LAE}. We find that LAEs are significantly more clustered  in quasar fields compared 
 with their clustering measured in blank fields. The fact that our measurements lie well above the \citet{Ouchi10} values
is again a manifestation of the overdensities of LAEs that we have uncovered around quasars. 

\subsection{Systematic Errors in our Measurements}
\label{sec:syst}

Our clustering analysis indicate that on the scales that we have probed ($R\lesssim7\,h^{-1}\,{\rm cMpc}$) we detect an enhancement of LAEs centered on quasars revealed by their spatial profile consistent with a power-law shape in the cross-correlation measurement (see Fig.~\ref{fig:crosscorr_LAE}). However,  the naive expectation of a deterministic bias model, whereby LAEs and quasars probe the same underlying dark mater overdensities, overpredicts the clustering of LAEs around quasars, with a correlation length $2.8$ times higher than the one measured in this work. Before discussing the implications of this discovery in section ~\S~\ref{sec:discussion}, it is important to establish
that it is not a result of systematic errors in our analysis.  There are several possible systematics which could
impact our results, which we discuss in turn.

First and foremost, are systematics associated with comparing LAEs selected
from one survey to the background number density determined from another
distinct survey. As discussed in section ~\S~\ref{ssec:selection}, the preferred
method would be to estimate the background directly from the same observations
used to conduct the clustering analysis. This is possible provided
either that  the field-of-view of the images is large enough that one
asymptotes to  the background level in the outermost regions
(i.e. where the clustering becomes vanishingly small), or
 if one has images with the same observational setup of blank fields. 
In both cases, the systematic errors associated with the background
number density
computation, coming from data reduction, source
detection (for example SExtractor parameters that were used),
photometry (e.g. aperture where colors are measured), LAE selection (i.e. color criteria, SExtractor flags, ${\rm S\slash N}$ ratio
of the selected sources, etc.), and photometric completeness computation drop out of the problem, since one
performs those steps in exactly the same manner for regions near quasars
as for the background, i.e. the clustering measurement is essentially a
relative measurement.

However, when the background level is computed from published
luminosity functions determined from a distinct blank field analysis,
these sources of systematics can creep in, because quasar fields could be
analyzed differently than the background fields. Our study potentially
suffers from this systematic, since the small field-of-view of FORS 
corresponding to $\sim 9.8\times9.8\,h^{-2}{\rm cMpc}^{2}$ precludes a background measurement
from our quasar fields, and we have no blank field pointings with our
observational setup.  In order to mitigate against this source of
systematic error, we carefully tailored our LAE selection to mimic the
selection adopted by \citet{Ouchi08} on which our background computation is
based, ensuring that we select LAEs with the same distribution of
equivalent widths and hence abundance, as we detailed in \S~\ref{ssec:selection} and \S~\ref{ssec:sample}. 

\begin{figure}[t!]
\begin{center}
\centering{\epsfig{file=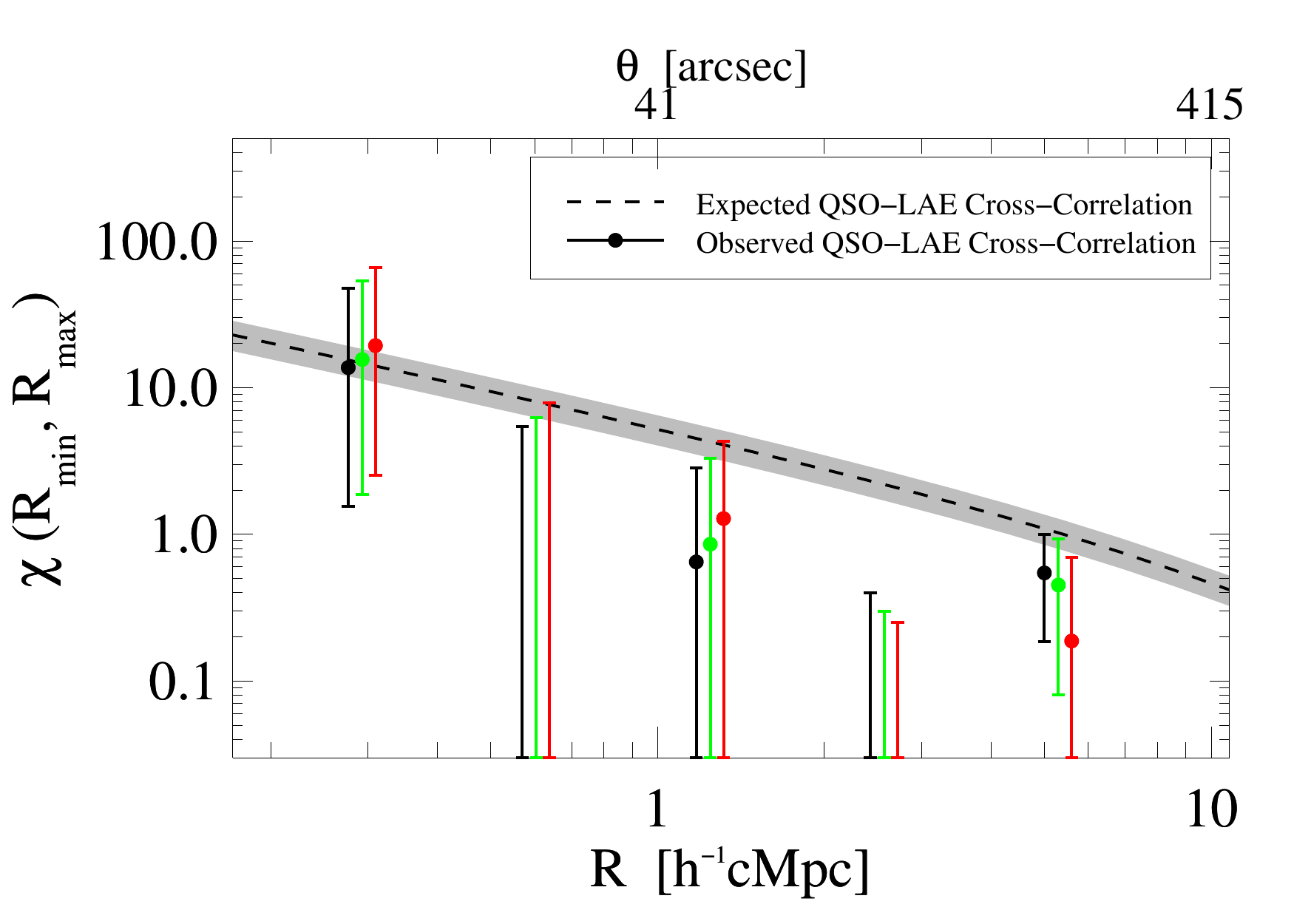, width=\columnwidth}}
\caption{Quasar-LAE cross-correlation function for three different LAEs samples. Black, green and red points show the cross-correlation function measured for a LAE sample $\lesssim10\%$, $\sim50\%$ and $\sim80\%$ complete respectively. We slightly offset the points on the x-axis for clarity. The dashed black line shows the theoretical expectation of $\chi$ for our 17 stacked fields computed from the quasar and LAEs auto-correlation functions and assuming a deterministic bias model (see \S~\ref{ssec:cross}). We obtain consistent clustering when we variate the limiting magnitude (and then the completeness correction) of the LAEs sample, which suggests that uncertainties in the completeness computation is not an important source of systematic errors in our measurement. \\}
\label{fig:test_comple}
\end{center}
\end{figure}

Another systematic error that could lead us to underestimate the LAE clustering is one associated with the computation of the photometric completeness. If we were significantly overestimating the completeness in the source detection, then the completeness-corrected number density in blank fields computed from \citet{Ouchi08}
would be overestimated and it would result in a much lower clustering. Note also that if the sample were highly complete, the computation of number counts in the background would not be highly sensitive to uncertainties in the completeness, however, if the sample were highly incomplete, the number counts would be highly sensitive to it. Given that the median completeness at our 5$\sigma$ flux limit
is only $<10$\% (see Fig.~\ref{fig:comple}), then when computing the blank field number density we are performing large corrections in the number counts predicted by the luminosity function measured by \citet{Ouchi08} which could be affected by large uncertainties. 
To address this possibility, and check that our large completeness correction is not strongly affecting our background estimation, we also performed our clustering analysis using two brighter LAE samples where the completeness is $\sim50$\% (corresponding to a median limiting magnitude HeI$\sim24.02$) and $\sim80$\% (corresponding to a median limiting magnitude HeI$\sim23.84$) which are composed of a total of 21 and 15 LAEs, respectively. When computing $n_{\rm G}$, we then integrated the luminosity function up to such brighter magnitudes, and then compute the quasar-LAE cross-correlation for both cases, which are shown in Fig. ~\ref{fig:test_comple}. If the completeness computation were inaccurate, then for the 50\% and 80\% complete samples, for which we avoid to perform large corrections, we would expect to obtain different clustering than the one computed in \S~\ref{ssec:cross} (for which the sample is $\lesssim10$\% complete). For both cases, we get similar results and verified that the parameter constraints do not change significantly which suggests that the completeness computation (and therefore the completeness-corrected number density $n_{\rm G}$) is not uncertain and that this is not a significant source of systematic error in this work.

Finally, an important issue is the accuracy of the quasar redshifts.
Large offsets between the quasar systemic redshift and the wavelengths
covered by our NB filter would imply that we are actually selecting
LAEs at sightly higher or lower redshifts, and hence at much larger
radial distances where the clustering signal would be vanishingly
small, such that the background number density of LAEs would be
expected. As described in 
\S~\ref{ssec:targets}, we computed the quasar systemic redshift by
using a custom line-centering code to determine the centroids of
emission lines (instead of using only the peaks of them), and used the
calibration of emission line shifts from \citet{Shen07}. Additionally,
we have selected only quasars with low redshift errors, and only
considered quasars for which the redshifted  Ly$\alpha$ line of LAEs would
land at the center of our NB filter (see Fig.~\ref{fig:qso_z}). All of these precautions considerably reduce
the possibility that redshift errors are biasing our analysis.
Also note that for three out of 17
quasars the redshift has been measured only using the CIV emission line
(considered to be the poorest redshift indicator), but for two of
those fields (SDSSJ0119--0342 and SDSSJ0850+0629) we have detected an
overdensity of LAEs greater than two, which indicates that we are
actually selecting LAEs in the quasar environment. In general, we do
not detect a correlation between the quasar redshift uncertainty
(reported in Table ~\ref{table:qso_prop}) and the amplitude of the
overdensity detected in their environment (reported in Table
~\ref{table:field_param}).

To summarize, we are confident that systematic errors associated with
the background determination, completeness corrections, and imprecise quasar
redshifts do not significantly impact our clustering measurements.

\section{Discussion}
\label{sec:discussion}

The enhancement of LAEs in quasar fields (by a factor of 1.4 compared
with the LAE number density in blank fields), the positive quasar-LAE
cross-correlation function (with correlation length
$r^{QG}_{0}=2.78^{+1.16}_{-1.05} \,h^{-1}\,{\rm cMpc}$ for a fixed
slope of $\gamma=1.8$), and the strong LAE auto-correlation function
(with an auto-correlation length 3.3 times higher than the LAE
auto-correlation length measured in blank fields) detected in this
work are all indicators that quasars trace massive dark matter
halos in the early universe. However, our results are in disagreement
with the expected quasar-LAE cross-correlation function computed
assuming a deterministic bias model, which overpredicts the
cross-correlation function by a factor of 6.4, as shown by the dashed line in
Fig.~\ref{fig:crosscorr_LAE}. In this section, we discuss some possible
explanations for this discrepancy.

The first possible explanation is that previous clustering measurements at $z\sim4$ (and used in this work) could suffer
from some systematic errors. If either the quasar clustering \citep[from][]{Shen07},
or the LAE clustering \citep[from][]{Ouchi10}), or both, were overestimated, we would
still expect an enhancement of LAEs in quasars fields over the background value, but
the expected quasar-LAE cross-correlation function computed in
section~\ref{sec:clustering} would be overestimated. Specifically, for
a deterministic bias model, where the cross-correlation function can be
written as $\xi_{QG} = \sqrt{ \xi_{QQ} \xi_{GG}}$, a discrepancy
of a factor of 6.4 between our measured cross-correlation function $\chi (R_{\rm min}, R_{\rm max})$
and the expected one can be explained by an overestimation
of the product $ \xi_{QQ} \xi_{GG}$ by a factor of $6.4^{2}= 40.96$,
implying that either the quasar or LAE correlation function (or both) would have to have been
highly overestimated\footnote{Note that given that $\chi (R_{\rm min}, R_{\rm max})$ is simply the integral of $\xi(R,Z)$ over the volume (see eqn.~(\ref{eq:estimator2_simpler})), an overprediction of $\chi (R_{\rm min}, R_{\rm max})$ by a factor of 6.4 ($\chi_{\rm exp} (R_{\rm min}, R_{\rm max})$/$\chi_{\rm obs} (R_{\rm min}, R_{\rm max}) = 6.4$) implies an overprediction of $\xi(R,Z)$ by the same factor ($\xi_{\rm exp} (R,Z)$/$\xi_{\rm obs} (R,Z) = 6.4$).}. An overestimate of this magnitude 
in the clustering of quasars or LAEs seems very unlikely and is highly inconsistent with the 1$\sigma$ quoted errors of previous works (represented by the gray shaded region in Fig.~\ref{fig:crosscorr_LAE}, which is computed by error propagation based on the reported $1\sigma$ errors of the $r^{QQ}_{0}$ and $r^{GG}_{0}$ parameters). Additionally, studies of high-redshift binary
  quasars, which provides an independent constraint on the quasar
  auto-correlation agree with the strong clustering reported in \citet{Shen07}
\citep{Hennawi10, McGreer16}.

Excluding the possibility of large errors in previous auto-correlation measurements, it seems
that the only explanation would be that the simple deterministic bias picture that we have assumed
breaks down.  This is in apparent contradiction of the recent work by \citep{garciavergara17} who measure the
cross-correlation of LBGs with quasars and found $r^{QG}_{0}=8.83^{+1.39}_{-1.51}\,h^{-1}\,{\rm Mpc}$ (for a fixed $\gamma=2.0$), which is actually in good agreement with the deterministic bias and the equation $\xi_{QG} = \sqrt{ \xi_{QQ} \xi_{GG}}$. 

Thus it seems that while LGBs are clustered around quasars in the right
numbers, it appears that LAEs may be avoiding quasar environments, at
least on scales of $\lesssim7\,h^{-1}$\,cMpc. Other searches for LAEs in high-redshift quasar fields at physical scales comparable to those probed by
our study also show that LAEs could be avoiding quasar environments, since the reported LAE number densities agrees
with the expectations from blank fields \citep{Banados13,Mazzucchelli17}. However conversely,
\citet{Swinbank12} detect a significant LAE overdensity in a $z=4.5$ quasar field. If we consider studies performed at
larger scales ($R\lesssim25\,h^{-1}$\,cMpc),  there are only few targeted LAEs in high-redshift ($z>4$) quasar environments. Some work
finds that on larger scales LAE overdensities are not present \citep{Kikuta17,Goto17,Ota18} or
that LAEs are distributed around the quasar but avoid their vicinity within $R<4.5$\,cMpc \citep{Kashikawa07}.
In summary,  studies of the clustering of LAEs around high-redshift quasars paint a confusing picture.
We note that all of the aforementioned work focused on  just one or two quasar fields, and considering the large
cosmic variance observed in our sample (see Table~\ref{table:field_param}), it is perhaps not surprising that the results
are not conclusive. More work is still required with much larger statistical samples.

It is possible that some extra piece of physics acting on the scales
probed in our work could be responsible for reducing the clustering of
LAEs around quasars, causing the deterministic bias picture to break
down. There are some theoretical ideas about how quasar feedback might
suppress star formation in the vicinity of a quasar
\citep[e.g][]{Francis04, Bruns12}, or how star formation in galaxies
located in dense fields could be affected via environmental quenching
\citep[e.g.][]{Peng10}. In both cases, we would expect a reduced
number of galaxies in our fields compared with the deterministic bias
model predictions, but the aforementioned physical processes would
need to be impacting neighboring galaxies at the relatively large
scales probed by our study ($\lesssim7\,h^{-1}$\,cMpc). However, it is
hard to understand why these processes would impact LAEs differently
from LBGs, which appear to obey deterministic bias.  Additionally, the
quasar feedback scenario seems inconsistent with the power-law shape
for the quasar-LAE cross-correlation detected in our work (see
Fig.~\ref{fig:crosscorr_LAE}), whereby LAEs tend to be preferentially
clustered (relative to the random expectation) very close to the
quasar (see also Fig.~\ref{fig:LAE_dist}).

One possible explanation for the detection of fewer LAEs in $z\sim4$
quasars environments might relate to the fact that the escape of
Ly$\alpha$ photons is particularly sensitive to the presence of
dust. If galaxies in the Mpc-scale quasar environments are on average
significantly more dusty, this could suppress the number of detected
LAEs, and explain why LBGs are less impacted.  Indeed, some studies of
quasar environments at $z\gtrsim4$ report the detection of close
submillimeter galaxies (SMGs) companions on scales of $R \lesssim
0.5$\,cMpc \citep[e.g.][]{Trakhtenbrot17,Decarli17,Bischetti18}, some
of which are totally extincted or at least very faint in the
rest-frame UV.  The acquisition of data at radio/submillimeter
wavelengths in our quasar fields would allow one to explore this
possibility, and would allow measurement of the
clustering properties of both optical and dusty galaxy populations
around quasars simultaneously. 

\section{Summary}
\label{sec:sum}
We studied the environment of 17 quasars at $z\sim4$ in order to make
the first measurement of the quasar-LAE cross-correlation function. We
imaged the quasar fields using a NB filter on VLT/FORS2, to select
LAEs at $z=3.88$. We carefully chose the quasar targets to have the
Ly$\alpha$ line located in the center of our NB filter and to have
small ($\rm 800\,km\,s^{-1}$) redshift uncertainties in order to ensure
that the selected LAEs are associated with the central quasar.

LAE selection was performed to match the criteria used by
\citet{Ouchi08} who selected LAEs at $z\sim4$ using the filter system
(V, NB570, B) to measure the LAE luminosity function.
Performing an analogous selection ensures
that we obtain an LAE sample with the same color completeness and
contamination levels. This allows us to simply use the
\citet{Ouchi08} luminosity
function to compute the blank field LAE number counts and perform a robust clustering analysis. 
We select LAEs to have an (rest-frame) EW$_{\rm Ly\alpha}>28$\AA\,
and find 25 in our fields. Given our survey volume and completeness, the number of LAEs expected in blank fields  is 18.36. Thus on average our work shows the quasar environment is
overdense by a factor of 1.4, indicating that
quasars inhabit massive dark matter halos in the young universe. Considering our fields
individually, we find that 10 out of 17 fields are underdense, whereas the
rest are overdense, demonstrating that
cosmic variance is large, and a significant source of noise in studies
of quasar environs. Studies at $z\sim 6$ often come to conclusions about quasar
environments based on studying a single or handfuls of quasar
fields, but the large cosmic variance we observe cautions one against over-interpreting
such small samples.

We measured the quasar-LAE cross-correlation function and find it is
consistent with a power-law shape, with correlation length of
$r^{QG}_{0}=2.78^{+1.16}_{-1.05} \,h^{-1}\,{\rm cMpc}$ for a fixed
slope of $\gamma=1.8$. For comparison, we also computed the expected
quasar-LAE cross-correlation function assuming a deterministic bias
model, and we find that this overpredicts the cross-correlation function by a factor of 6.4 in comparison with what is measured in this work, 
 or equivalently, a factor of 2.1 in the overall number of galaxies. We also measure the LAE
auto-correlation in quasar fields and find a correlation length of
$r^{GG}_{0} = 9.12^{+1.32}_{-1.31}\,h^{-1}$\,cMpc which is $3.3$ times
higher than the LAE correlation length measured in blank fields, providing
further confirmation that quasars are indeed
tracing biased regions of the universe.

We discussed possible reasons why the deterministic bias picture could break down for LAEs,
although it is puzzling that it appears to hold for LBGs. If feedback or
something specific to the quasar environment is responsible for preferentially
suppressing LAE clustering around quasars, then this effect ought to only
act within some ``sphere of influence'' around the quasar. Hence a larger scale
study of the clustering of both LAE and LBGs that extends beyond the scales probed here
($R\lesssim7\,h^{-1}\,{\rm cMpc}$) would be particularly illuminating. It may also be that LAEs are absent in the
quasar environment because galaxies near quasars  are on average more dusty. Future radio and submillimeter studies
of these quasar fields would allow one to measure the clustering
of both radio/submillimeter and optical selected galaxy populations simultaneously, which is an important
avenue for future work.

\acknowledgments
We acknowledge Gabor Worseck, Yue Shen, Arjen van der Wel, Bram Venemans, and Masami Ouchi for kindly providing useful data and material used in this paper. We thank Jacqueline Hodge for her support during the final phase of the project. CGV acknowledges the support from CONICYT Doctoral Fellowship Programme (CONICYT-PCHA/Doctorado Nacional 2012-21120442), DAAD in the context of the PUC-HD Graduate Exchange Fellowship, and CONICYT Postdoctoral Fellowship Programme (CONICYT-PCHA/Postdoctorado Becas Chile 2017-74180018). LFB acknowledges the support from CONICYT Project BASAL AFB- 170002.

\bibliography{my_bib}

\begin{thebibliography}{}
\expandafter\ifx\csname natexlab\endcsname\relax\def\natexlab#1{#1}\fi

\bibitem[{{Adams} {et~al.}(2015){Adams}, {Martini}, {Croxall}, {Overzier}, \&
  {Silverman}}]{Adams15}
{Adams}, S.~M., {Martini}, P., {Croxall}, K.~V., {Overzier}, R.~A., \&
  {Silverman}, J.~D. 2015, \mnras, 448, 1335

\bibitem[{{Adelberger} \& {Steidel}(2005)}]{Adelberger05}
{Adelberger}, K.~L., \& {Steidel}, C.~C. 2005, \apj, 630, 50

\bibitem[{{Angulo} {et~al.}(2012){Angulo}, {Springel}, {White}, {Cole},
  {Jenkins}, {Baugh}, \& {Frenk}}]{angulo12}
{Angulo}, R.~E., {Springel}, V., {White}, S.~D.~M., {et~al.} 2012, \mnras, 425,
  2722

\bibitem[{{Appenzeller} \& {Rupprecht}(1992)}]{Appenzeller92}
{Appenzeller}, I., \& {Rupprecht}, G. 1992, The Messenger, 67, 18

\bibitem[{{Ba{\~n}ados} {et~al.}(2013){Ba{\~n}ados}, {Venemans}, {Walter},
  {Kurk}, {Overzier}, \& {Ouchi}}]{Banados13}
{Ba{\~n}ados}, E., {Venemans}, B., {Walter}, F., {et~al.} 2013, \apj, 773, 178

\bibitem[{{Balmaverde} {et~al.}(2017){Balmaverde}, {Gilli}, {Mignoli},
  {Bolzonella}, {Brusa}, {Cappelluti}, {Comastri}, {Sani}, {Vanzella},
  {Vignali}, {Vito}, \& {Zamorani}}]{Balmaverde17}
{Balmaverde}, B., {Gilli}, R., {Mignoli}, M., {et~al.} 2017, \aap, 606, A23

\bibitem[{{Becker} {et~al.}(1995){Becker}, {White}, \& {Helfand}}]{Becker95}
{Becker}, R.~H., {White}, R.~L., \& {Helfand}, D.~J. 1995, \apj, 450, 559

\bibitem[{{Bertin}(2006)}]{Bertin06}
{Bertin}, E. 2006, in Astronomical Society of the Pacific Conference Series,
  Vol. 351, Astronomical Data Analysis Software and Systems XV, ed.
  C.~{Gabriel}, C.~{Arviset}, D.~{Ponz}, \& S.~{Enrique}, 112

\bibitem[{{Bertin} \& {Arnouts}(1996)}]{Bertin96}
{Bertin}, E., \& {Arnouts}, S. 1996, \aaps, 117, 393

\bibitem[{{Bertin} {et~al.}(2002){Bertin}, {Mellier}, {Radovich}, {Missonnier},
  {Didelon}, \& {Morin}}]{Bertin02}
{Bertin}, E., {Mellier}, Y., {Radovich}, M., {et~al.} 2002, in Astronomical
  Society of the Pacific Conference Series, Vol. 281, Astronomical Data
  Analysis Software and Systems XI, ed. D.~A. {Bohlender}, D.~{Durand}, \&
  T.~H. {Handley}, 228

\bibitem[{{Bischetti} {et~al.}(2018){Bischetti}, {Piconcelli}, {Feruglio},
  {Duras}, {Bongiorno}, {Carniani}, {Marconi}, {Pappalardo}, {Schneider},
  {Travascio}, {Valiante}, {Vietri}, {Zappacosta}, \& {Fiore}}]{Bischetti18}
{Bischetti}, M., {Piconcelli}, E., {Feruglio}, C., {et~al.} 2018, \aap, 617,
  A82

\bibitem[{{Bouwens} {et~al.}(2007){Bouwens}, {Illingworth}, {Franx}, \&
  {Ford}}]{Bouwens07}
{Bouwens}, R.~J., {Illingworth}, G.~D., {Franx}, M., \& {Ford}, H. 2007, \apj,
  670, 928

\bibitem[{{Bouwens} {et~al.}(2009){Bouwens}, {Illingworth}, {Franx}, {Chary},
  {Meurer}, {Conselice}, {Ford}, {Giavalisco}, \& {van Dokkum}}]{Bouwens09}
{Bouwens}, R.~J., {Illingworth}, G.~D., {Franx}, M., {et~al.} 2009, \apj, 705,
  936

\bibitem[{{Bouwens} {et~al.}(2010){Bouwens}, {Illingworth}, {Oesch},
  {Stiavelli}, {van Dokkum}, {Trenti}, {Magee}, {Labb{\'e}}, {Franx},
  {Carollo}, \& {Gonzalez}}]{Bouwens10}
{Bouwens}, R.~J., {Illingworth}, G.~D., {Oesch}, P.~A., {et~al.} 2010, \apjl,
  709, L133

\bibitem[{{Brammer} {et~al.}(2008){Brammer}, {van Dokkum}, \&
  {Coppi}}]{Brammer08}
{Brammer}, G.~B., {van Dokkum}, P.~G., \& {Coppi}, P. 2008, \apj, 686, 1503

\bibitem[{{Bruns} {et~al.}(2012){Bruns}, {Wyithe}, {Bland-Hawthorn}, \&
  {Dijkstra}}]{Bruns12}
{Bruns}, L.~R., {Wyithe}, J.~S.~B., {Bland-Hawthorn}, J., \& {Dijkstra}, M.
  2012, \mnras, 421, 2543

\bibitem[{{Bruzual} \& {Charlot}(2003)}]{Bruzual03}
{Bruzual}, G., \& {Charlot}, S. 2003, \mnras, 344, 1000

\bibitem[{{Capak} {et~al.}(2011){Capak}, {Riechers}, {Scoville}, {Carilli},
  {Cox}, {Neri}, {Robertson}, {Salvato}, {Schinnerer}, {Yan}, {Wilson}, {Yun},
  {Civano}, {Elvis}, {Karim}, {Mobasher}, \& {Staguhn}}]{Capak11}
{Capak}, P.~L., {Riechers}, D., {Scoville}, N.~Z., {et~al.} 2011, \nat, 470,
  233

\bibitem[{{Cardelli} {et~al.}(1989){Cardelli}, {Clayton}, \&
  {Mathis}}]{Cardelli89}
{Cardelli}, J.~A., {Clayton}, G.~C., \& {Mathis}, J.~S. 1989, \apj, 345, 245

\bibitem[{{Cassata} {et~al.}(2011){Cassata}, {Le F{\`e}vre}, {Garilli},
  {Maccagni}, {Le Brun}, {Scodeggio}, {Tresse}, {Ilbert}, {Zamorani},
  {Cucciati}, {Contini}, {Bielby}, {Mellier}, {McCracken}, {Pollo},
  {Zanichelli}, {Bardelli}, {Cappi}, {Pozzetti}, {Vergani}, \&
  {Zucca}}]{Cassata11}
{Cassata}, P., {Le F{\`e}vre}, O., {Garilli}, B., {et~al.} 2011, \aap, 525,
  A143

\bibitem[{{Chabrier}(2003)}]{Chabrier03}
{Chabrier}, G. 2003, \pasp, 115, 763

\bibitem[{{Chiang} {et~al.}(2013){Chiang}, {Overzier}, \&
  {Gebhardt}}]{Chiang13}
{Chiang}, Y.-K., {Overzier}, R., \& {Gebhardt}, K. 2013, \apj, 779, 127

\bibitem[{{Coatman} {et~al.}(2017){Coatman}, {Hewett}, {Banerji}, {Richards},
  {Hennawi}, \& {Prochaska}}]{Coatman17}
{Coatman}, L., {Hewett}, P.~C., {Banerji}, M., {et~al.} 2017, \mnras, 465, 2120

\bibitem[{{Coil} {et~al.}(2007){Coil}, {Hennawi}, {Newman}, {Cooper}, \&
  {Davis}}]{Coil07}
{Coil}, A.~L., {Hennawi}, J.~F., {Newman}, J.~A., {Cooper}, M.~C., \& {Davis},
  M. 2007, \apj, 654, 115

\bibitem[{{Dawson} {et~al.}(2013){Dawson}, {Schlegel}, {Ahn}, {Anderson},
  {Aubourg}, {Bailey}, {Barkhouser}, {Bautista}, {Beifiori}, {Berlind},
  {Bhardwaj}, {Bizyaev}, {Blake}, {Blanton}, {Blomqvist}, {Bolton}, {Borde},
  {Bovy}, {Brandt}, {Brewington}, {Brinkmann}, {Brown}, {Brownstein}, {Bundy},
  {Busca}, {Carithers}, {Carnero}, {Carr}, {Chen}, {Comparat}, {Connolly},
  {Cope}, {Croft}, {Cuesta}, {da Costa}, {Davenport}, {Delubac}, {de Putter},
  {Dhital}, {Ealet}, {Ebelke}, {Eisenstein}, {Escoffier}, {Fan}, {Filiz Ak},
  {Finley}, {Font-Ribera}, {G{\'e}nova-Santos}, {Gunn}, {Guo}, {Haggard},
  {Hall}, {Hamilton}, {Harris}, {Harris}, {Ho}, {Hogg}, {Holder}, {Honscheid},
  {Huehnerhoff}, {Jordan}, {Jordan}, {Kauffmann}, {Kazin}, {Kirkby}, {Klaene},
  {Kneib}, {Le Goff}, {Lee}, {Long}, {Loomis}, {Lundgren}, {Lupton}, {Maia},
  {Makler}, {Malanushenko}, {Malanushenko}, {Mandelbaum}, {Manera}, {Maraston},
  {Margala}, {Masters}, {McBride}, {McDonald}, {McGreer}, {McMahon}, {Mena},
  {Miralda-Escud{\'e}}, {Montero-Dorta}, {Montesano}, {Muna}, {Myers},
  {Naugle}, {Nichol}, {Noterdaeme}, {Nuza}, {Olmstead}, {Oravetz}, {Oravetz},
  {Owen}, {Padmanabhan}, {Palanque-Delabrouille}, {Pan}, {Parejko},
  {P{\^a}ris}, {Percival}, {P{\'e}rez-Fournon}, {P{\'e}rez-R{\`a}fols},
  {Petitjean}, {Pfaffenberger}, {Pforr}, {Pieri}, {Prada}, {Price-Whelan},
  {Raddick}, {Rebolo}, {Rich}, {Richards}, {Rockosi}, {Roe}, {Ross}, {Ross},
  {Rossi}, {Rubi{\~n}o-Martin}, {Samushia}, {S{\'a}nchez}, {Sayres}, {Schmidt},
  {Schneider}, {Sc{\'o}ccola}, {Seo}, {Shelden}, {Sheldon}, {Shen}, {Shu},
  {Slosar}, {Smee}, {Snedden}, {Stauffer}, {Steele}, {Strauss}, {Streblyanska},
  {Suzuki}, {Swanson}, {Tal}, {Tanaka}, {Thomas}, {Tinker}, {Tojeiro},
  {Tremonti}, {Vargas Maga{\~n}a}, {Verde}, {Viel}, {Wake}, {Watson}, {Weaver},
  {Weinberg}, {Weiner}, {West}, {White}, {Wood-Vasey}, {Yeche}, {Zehavi},
  {Zhao}, \& {Zheng}}]{Dawson13}
{Dawson}, K.~S., {Schlegel}, D.~J., {Ahn}, C.~P., {et~al.} 2013, \aj, 145, 10

\bibitem[{{Decarli} {et~al.}(2017){Decarli}, {Walter}, {Venemans},
  {Ba{\~n}ados}, {Bertoldi}, {Carilli}, {Fan}, {Farina}, {Mazzucchelli},
  {Riechers}, {Rix}, {Strauss}, {Wang}, \& {Yang}}]{Decarli17}
{Decarli}, R., {Walter}, F., {Venemans}, B.~P., {et~al.} 2017, \nat, 545, 457

\bibitem[{{Drake} {et~al.}(2017){Drake}, {Garel}, {Wisotzki}, {Leclercq},
  {Hashimoto}, {Richard}, {Bacon}, {Blaizot}, {Caruana}, {Conseil}, {Contini},
  {Guiderdoni}, {Herenz}, {Inami}, {Lewis}, {Mahler}, {Marino}, {Pello},
  {Schaye}, {Verhamme}, {Ventou}, \& {Weilbacher}}]{Drake17}
{Drake}, A.~B., {Garel}, T., {Wisotzki}, L., {et~al.} 2017, \aap, 608, A6

\bibitem[{{Eftekharzadeh} {et~al.}(2015){Eftekharzadeh}, {Myers}, {White},
  {Weinberg}, {Schneider}, {Shen}, {Font-Ribera}, {Ross}, {Paris}, \&
  {Streblyanska}}]{Eftekharzadeh15}
{Eftekharzadeh}, S., {Myers}, A.~D., {White}, M., {et~al.} 2015, \mnras, 453,
  2779

\bibitem[{{Eisenstein} {et~al.}(2011){Eisenstein}, {Weinberg}, {Agol},
  {Aihara}, {Allende Prieto}, {Anderson}, {Arns}, {Aubourg}, {Bailey},
  {Balbinot}, \& et~al.}]{Eisenstein11}
{Eisenstein}, D.~J., {Weinberg}, D.~H., {Agol}, E., {et~al.} 2011, \aj, 142, 72

\bibitem[{{Ferrarese} \& {Merritt}(2000)}]{Ferrarese00}
{Ferrarese}, L., \& {Merritt}, D. 2000, \apjl, 539, L9

\bibitem[{{Francis} \& {Bland-Hawthorn}(2004)}]{Francis04}
{Francis}, P.~J., \& {Bland-Hawthorn}, J. 2004, \mnras, 353, 301

\bibitem[{{Fukugita} {et~al.}(1995){Fukugita}, {Shimasaku}, \&
  {Ichikawa}}]{Fukugita95}
{Fukugita}, M., {Shimasaku}, K., \& {Ichikawa}, T. 1995, \pasp, 107, 945

\bibitem[{{Garc{\'{\i}}a-Vergara} {et~al.}(2017){Garc{\'{\i}}a-Vergara},
  {Hennawi}, {Barrientos}, \& {Rix}}]{garciavergara17}
{Garc{\'{\i}}a-Vergara}, C., {Hennawi}, J.~F., {Barrientos}, L.~F., \& {Rix},
  H.-W. 2017, \apj, 848, 7

\bibitem[{{Gaskell}(1982)}]{Gaskell82}
{Gaskell}, C.~M. 1982, \apj, 263, 79

\bibitem[{{Gebhardt} {et~al.}(2000){Gebhardt}, {Bender}, {Bower}, {Dressler},
  {Faber}, {Filippenko}, {Green}, {Grillmair}, {Ho}, {Kormendy}, {Lauer},
  {Magorrian}, {Pinkney}, {Richstone}, \& {Tremaine}}]{Gebhardt00}
{Gebhardt}, K., {Bender}, R., {Bower}, G., {et~al.} 2000, \apjl, 539, L13

\bibitem[{{Gehrels}(1986)}]{Gehrels86}
{Gehrels}, N. 1986, \apj, 303, 336

\bibitem[{{Goto} {et~al.}(2017){Goto}, {Utsumi}, {Kikuta}, {Miyazaki}, {Shiki},
  \& {Hashimoto}}]{Goto17}
{Goto}, T., {Utsumi}, Y., {Kikuta}, S., {et~al.} 2017, \mnras, 470, L117

\bibitem[{{Hamuy} {et~al.}(1994){Hamuy}, {Suntzeff}, {Heathcote}, {Walker},
  {Gigoux}, \& {Phillips}}]{Hamuy94}
{Hamuy}, M., {Suntzeff}, N.~B., {Heathcote}, S.~R., {et~al.} 1994, \pasp, 106,
  566

\bibitem[{{Hamuy} {et~al.}(1992){Hamuy}, {Walker}, {Suntzeff}, {Gigoux},
  {Heathcote}, \& {Phillips}}]{Hamuy92}
{Hamuy}, M., {Walker}, A.~R., {Suntzeff}, N.~B., {et~al.} 1992, \pasp, 104, 533

\bibitem[{{He} {et~al.}(2018){He}, {Akiyama}, {Bosch}, {Enoki}, {Harikane},
  {Ikeda}, {Kashikawa}, {Kawaguchi}, {Komiyama}, {Lee}, {Matsuoka}, {Miyazaki},
  {Nagao}, {Nagashima}, {Niida}, {Nishizawa}, {Oguri}, {Onoue}, {Oogi},
  {Ouchi}, {Schulze}, {Shirasaki}, {Silverman}, {Tanaka}, {Tanaka}, {Toba},
  {Uchiyama}, \& {Yamashita}}]{He18}
{He}, W., {Akiyama}, M., {Bosch}, J., {et~al.} 2018, \pasj, 70, S33

\bibitem[{{Hennawi} {et~al.}(2006){Hennawi}, {Prochaska}, {Burles}, {Strauss},
  {Richards}, {Schlegel}, {Fan}, {Schneider}, {Zakamska}, {Oguri}, {Gunn},
  {Lupton}, \& {Brinkmann}}]{hennawi06b}
{Hennawi}, J.~F., {Prochaska}, J.~X., {Burles}, S., {et~al.} 2006, \apj, 651,
  61

\bibitem[{{Hennawi} {et~al.}(2010){Hennawi}, {Myers}, {Shen}, {Strauss},
  {Djorgovski}, {Fan}, {Glikman}, {Mahabal}, {Martin}, {Richards}, {Schneider},
  \& {Shankar}}]{Hennawi10}
{Hennawi}, J.~F., {Myers}, A.~D., {Shen}, Y., {et~al.} 2010, \apj, 719, 1672

\bibitem[{{Husband} {et~al.}(2013){Husband}, {Bremer}, {Stanway}, {Davies},
  {Lehnert}, \& {Douglas}}]{Husband13}
{Husband}, K., {Bremer}, M.~N., {Stanway}, E.~R., {et~al.} 2013, \mnras, 432,
  2869

\bibitem[{{Ikeda} {et~al.}(2015){Ikeda}, {Nagao}, {Taniguchi}, {Matsuoka},
  {Kajisawa}, {Akiyama}, {Miyaji}, {Kashikawa}, {Morokuma}, {Shioya}, {Enoki},
  {Capak}, {Koekemoer}, {Masters}, {Salvato}, {Sanders}, {Schinnerer}, \&
  {Scoville}}]{Ikeda15}
{Ikeda}, H., {Nagao}, T., {Taniguchi}, Y., {et~al.} 2015, \apj, 809, 138

\bibitem[{{Kashikawa} {et~al.}(2007){Kashikawa}, {Kitayama}, {Doi}, {Misawa},
  {Komiyama}, \& {Ota}}]{Kashikawa07}
{Kashikawa}, N., {Kitayama}, T., {Doi}, M., {et~al.} 2007, \apj, 663, 765

\bibitem[{{Kashikawa} {et~al.}(2006){Kashikawa}, {Yoshida}, {Shimasaku},
  {Nagashima}, {Yahagi}, {Ouchi}, {Matsuda}, {Malkan}, {Doi}, {Iye}, {Ajiki},
  {Akiyama}, {Ando}, {Aoki}, {Furusawa}, {Hayashino}, {Iwamuro}, {Karoji},
  {Kobayashi}, {Kodaira}, {Kodama}, {Komiyama}, {Miyazaki}, {Mizumoto},
  {Morokuma}, {Motohara}, {Murayama}, {Nagao}, {Nariai}, {Ohta}, {Okamura},
  {Sasaki}, {Sato}, {Sekiguchi}, {Shioya}, {Tamura}, {Taniguchi}, {Umemura},
  {Yamada}, \& {Yasuda}}]{Kashikawa06}
{Kashikawa}, N., {Yoshida}, M., {Shimasaku}, K., {et~al.} 2006, \apj, 637, 631

\bibitem[{{Kikuta} {et~al.}(2017){Kikuta}, {Imanishi}, {Matsuoka}, {Matsuda},
  {Shimasaku}, \& {Nakata}}]{Kikuta17}
{Kikuta}, S., {Imanishi}, M., {Matsuoka}, Y., {et~al.} 2017, \apj, 841, 128

\bibitem[{{Kim} {et~al.}(2009){Kim}, {Stiavelli}, {Trenti}, {Pavlovsky},
  {Djorgovski}, {Scarlata}, {Stern}, {Mahabal}, {Thompson}, {Dickinson},
  {Panagia}, \& {Meylan}}]{Kim09}
{Kim}, S., {Stiavelli}, M., {Trenti}, M., {et~al.} 2009, \apj, 695, 809

\bibitem[{{Madau}(1995)}]{Madau95}
{Madau}, P. 1995, \apj, 441, 18

\bibitem[{{Magorrian} {et~al.}(1998){Magorrian}, {Tremaine}, {Richstone},
  {Bender}, {Bower}, {Dressler}, {Faber}, {Gebhardt}, {Green}, {Grillmair},
  {Kormendy}, \& {Lauer}}]{Magorrian98}
{Magorrian}, J., {Tremaine}, S., {Richstone}, D., {et~al.} 1998, \aj, 115, 2285

\bibitem[{{Mazzucchelli} {et~al.}(2017){Mazzucchelli}, {Ba{\~n}ados},
  {Decarli}, {Farina}, {Venemans}, {Walter}, \& {Overzier}}]{Mazzucchelli17}
{Mazzucchelli}, C., {Ba{\~n}ados}, E., {Decarli}, R., {et~al.} 2017, \apj, 834,
  83

\bibitem[{{McGreer} {et~al.}(2016){McGreer}, {Eftekharzadeh}, {Myers}, \&
  {Fan}}]{McGreer16}
{McGreer}, I.~D., {Eftekharzadeh}, S., {Myers}, A.~D., \& {Fan}, X. 2016, \aj,
  151, 61

\bibitem[{{Metcalfe} {et~al.}(2001){Metcalfe}, {Shanks}, {Campos}, {McCracken},
  \& {Fong}}]{Metcalfe01}
{Metcalfe}, N., {Shanks}, T., {Campos}, A., {McCracken}, H.~J., \& {Fong}, R.
  2001, \mnras, 323, 795

\bibitem[{{Mo} \& {White}(1996)}]{Mo96}
{Mo}, H.~J., \& {White}, S.~D.~M. 1996, \mnras, 282, 347

\bibitem[{{Morselli} {et~al.}(2014){Morselli}, {Mignoli}, {Gilli}, {Vignali},
  {Comastri}, {Sani}, {Cappelluti}, {Zamorani}, {Brusa}, {Gallozzi}, \&
  {Vanzella}}]{Morselli14}
{Morselli}, L., {Mignoli}, M., {Gilli}, R., {et~al.} 2014, \aap, 568, A1

\bibitem[{{Myers} {et~al.}(2006){Myers}, {Brunner}, {Richards}, {Nichol},
  {Schneider}, {Vanden Berk}, {Scranton}, {Gray}, \& {Brinkmann}}]{Myers06}
{Myers}, A.~D., {Brunner}, R.~J., {Richards}, G.~T., {et~al.} 2006, \apj, 638,
  622

\bibitem[{{Oke}(1974)}]{Oke74}
{Oke}, J.~B. 1974, \apjs, 27, 21

\bibitem[{{Oke}(1990)}]{Oke90}
---. 1990, \aj, 99, 1621

\bibitem[{{Ota} {et~al.}(2018){Ota}, {Venemans}, {Taniguchi}, {Kashikawa},
  {Nakata}, {Harikane}, {Ba{\~n}ados}, {Overzier}, {Riechers}, {Walter},
  {Toshikawa}, {Shibuya}, \& {Jiang}}]{Ota18}
{Ota}, K., {Venemans}, B.~P., {Taniguchi}, Y., {et~al.} 2018, \apj, 856, 109

\bibitem[{{Ouchi} {et~al.}(2004{\natexlab{a}}){Ouchi}, {Shimasaku}, {Okamura},
  {Furusawa}, {Kashikawa}, {Ota}, {Doi}, {Hamabe}, {Kimura}, {Komiyama},
  {Miyazaki}, {Miyazaki}, {Nakata}, {Sekiguchi}, {Yagi}, \&
  {Yasuda}}]{Ouchi04a}
{Ouchi}, M., {Shimasaku}, K., {Okamura}, S., {et~al.} 2004{\natexlab{a}}, \apj,
  611, 660

\bibitem[{{Ouchi} {et~al.}(2004{\natexlab{b}}){Ouchi}, {Shimasaku}, {Okamura},
  {Furusawa}, {Kashikawa}, {Ota}, {Doi}, {Hamabe}, {Kimura}, {Komiyama},
  {Miyazaki}, {Miyazaki}, {Nakata}, {Sekiguchi}, {Yagi}, \&
  {Yasuda}}]{Ouchi04b}
---. 2004{\natexlab{b}}, \apj, 611, 685

\bibitem[{{Ouchi} {et~al.}(2008){Ouchi}, {Shimasaku}, {Akiyama}, {Simpson},
  {Saito}, {Ueda}, {Furusawa}, {Sekiguchi}, {Yamada}, {Kodama}, {Kashikawa},
  {Okamura}, {Iye}, {Takata}, {Yoshida}, \& {Yoshida}}]{Ouchi08}
{Ouchi}, M., {Shimasaku}, K., {Akiyama}, M., {et~al.} 2008, \apjs, 176, 301

\bibitem[{{Ouchi} {et~al.}(2010){Ouchi}, {Shimasaku}, {Furusawa}, {Saito},
  {Yoshida}, {Akiyama}, {Ono}, {Yamada}, {Ota}, {Kashikawa}, {Iye}, {Kodama},
  {Okamura}, {Simpson}, \& {Yoshida}}]{Ouchi10}
{Ouchi}, M., {Shimasaku}, K., {Furusawa}, H., {et~al.} 2010, \apj, 723, 869

\bibitem[{{Ouchi} {et~al.}(2018){Ouchi}, {Harikane}, {Shibuya}, {Shimasaku},
  {Taniguchi}, {Konno}, {Kobayashi}, {Kajisawa}, {Nagao}, {Ono}, {Inoue},
  {Umemura}, {Mori}, {Hasegawa}, {Higuchi}, {Komiyama}, {Matsuda}, {Nakajima},
  {Saito}, \& {Wang}}]{Ouchi18}
{Ouchi}, M., {Harikane}, Y., {Shibuya}, T., {et~al.} 2018, \pasj, 70, S13

\bibitem[{{Overzier} {et~al.}(2008){Overzier}, {Bouwens}, {Cross}, {Venemans},
  {Miley}, {Zirm}, {Ben{\'{\i}}tez}, {Blakeslee}, {Coe}, {Demarco}, {Ford},
  {Homeier}, {Illingworth}, {Kurk}, {Martel}, {Mei}, {Oliveira},
  {R{\"o}ttgering}, {Tsvetanov}, \& {Zheng}}]{Overzier08}
{Overzier}, R.~A., {Bouwens}, R.~J., {Cross}, N.~J.~G., {et~al.} 2008, \apj,
  673, 143

\bibitem[{{Padmanabhan} {et~al.}(2009){Padmanabhan}, {White}, {Norberg}, \&
  {Porciani}}]{Padmanabhan09}
{Padmanabhan}, N., {White}, M., {Norberg}, P., \& {Porciani}, C. 2009, \mnras,
  397, 1862

\bibitem[{{P{\^a}ris} {et~al.}(2014){P{\^a}ris}, {Petitjean}, {Aubourg},
  {Ross}, {Myers}, {Streblyanska}, {Bailey}, {Hall}, {Strauss}, {Anderson},
  {Bizyaev}, {Borde}, {Brinkmann}, {Bovy}, {Brandt}, {Brewington},
  {Brownstein}, {Cook}, {Ebelke}, {Fan}, {Filiz Ak}, {Finley}, {Font-Ribera},
  {Ge}, {Hamann}, {Ho}, {Jiang}, {Kinemuchi}, {Malanushenko}, {Malanushenko},
  {Marchante}, {McGreer}, {McMahon}, {Miralda-Escud{\'e}}, {Muna},
  {Noterdaeme}, {Oravetz}, {Palanque-Delabrouille}, {Pan}, {Perez-Fournon},
  {Pieri}, {Riffel}, {Schlegel}, {Schneider}, {Simmons}, {Viel}, {Weaver},
  {Wood-Vasey}, {Y{\`e}che}, \& {York}}]{Paris14}
{P{\^a}ris}, I., {Petitjean}, P., {Aubourg}, {\'E}., {et~al.} 2014, \aap, 563,
  A54

\bibitem[{{Patat} {et~al.}(2011){Patat}, {Moehler}, {O'Brien}, {Pompei},
  {Bensby}, {Carraro}, {de Ugarte Postigo}, {Fox}, {Gavignaud}, {James},
  {Korhonen}, {Ledoux}, {Randall}, {Sana}, {Smoker}, {Stefl}, \&
  {Szeifert}}]{Patat11}
{Patat}, F., {Moehler}, S., {O'Brien}, K., {et~al.} 2011, \aap, 527, A91

\bibitem[{{Peng} {et~al.}(2010){Peng}, {Lilly}, {Kova{\v c}}, {Bolzonella},
  {Pozzetti}, {Renzini}, {Zamorani}, {Ilbert}, {Knobel}, {Iovino}, {Maier},
  {Cucciati}, {Tasca}, {Carollo}, {Silverman}, {Kampczyk}, {de Ravel},
  {Sanders}, {Scoville}, {Contini}, {Mainieri}, {Scodeggio}, {Kneib}, {Le
  F{\`e}vre}, {Bardelli}, {Bongiorno}, {Caputi}, {Coppa}, {de la Torre},
  {Franzetti}, {Garilli}, {Lamareille}, {Le Borgne}, {Le Brun}, {Mignoli},
  {Perez Montero}, {Pello}, {Ricciardelli}, {Tanaka}, {Tresse}, {Vergani},
  {Welikala}, {Zucca}, {Oesch}, {Abbas}, {Barnes}, {Bordoloi}, {Bottini},
  {Cappi}, {Cassata}, {Cimatti}, {Fumana}, {Hasinger}, {Koekemoer},
  {Leauthaud}, {Maccagni}, {Marinoni}, {McCracken}, {Memeo}, {Meneux}, {Nair},
  {Porciani}, {Presotto}, \& {Scaramella}}]{Peng10}
{Peng}, Y.-j., {Lilly}, S.~J., {Kova{\v c}}, K., {et~al.} 2010, \apj, 721, 193

\bibitem[{{Planck Collaboration} {et~al.}(2018){Planck Collaboration},
  {Aghanim}, {Akrami}, {Ashdown}, {Aumont}, {Baccigalupi}, {Ballardini},
  {Banday}, {Barreiro}, {Bartolo}, {Basak}, {Battye}, {Benabed}, {Bernard},
  {Bersanelli}, {Bielewicz}, {Bock}, {Bond}, {Borrill}, {Bouchet}, {Boulanger},
  {Bucher}, {Burigana}, {Butler}, {Calabrese}, {Cardoso}, {Carron},
  {Challinor}, {Chiang}, {Chluba}, {Colombo}, {Combet}, {Contreras}, {Crill},
  {Cuttaia}, {de Bernardis}, {de Zotti}, {Delabrouille}, {Delouis}, {Di
  Valentino}, {Diego}, {Dor{\'e}}, {Douspis}, {Ducout}, {Dupac}, {Dusini},
  {Efstathiou}, {Elsner}, {En{\ss}lin}, {Eriksen}, {Fantaye}, {Farhang},
  {Fergusson}, {Fernandez-Cobos}, {Finelli}, {Forastieri}, {Frailis},
  {Franceschi}, {Frolov}, {Galeotta}, {Galli}, {Ganga}, {G{\'e}nova-Santos},
  {Gerbino}, {Ghosh}, {Gonz{\'a}lez-Nuevo}, {G{\'o}rski}, {Gratton},
  {Gruppuso}, {Gudmundsson}, {Hamann}, {Handley}, {Herranz}, {Hivon}, {Huang},
  {Jaffe}, {Jones}, {Karakci}, {Keih{\"a}nen}, {Keskitalo}, {Kiiveri}, {Kim},
  {Kisner}, {Knox}, {Krachmalnicoff}, {Kunz}, {Kurki-Suonio}, {Lagache},
  {Lamarre}, {Lasenby}, {Lattanzi}, {Lawrence}, {Le Jeune}, {Lemos},
  {Lesgourgues}, {Levrier}, {Lewis}, {Liguori}, {Lilje}, {Lilley}, {Lindholm},
  {L{\'o}pez-Caniego}, {Lubin}, {Ma}, {Mac{\'{\i}}as-P{\'e}rez}, {Maggio},
  {Maino}, {Mandolesi}, {Mangilli}, {Marcos-Caballero}, {Maris}, {Martin},
  {Martinelli}, {Mart{\'{\i}}nez-Gonz{\'a}lez}, {Matarrese}, {Mauri}, {McEwen},
  {Meinhold}, {Melchiorri}, {Mennella}, {Migliaccio}, {Millea}, {Mitra},
  {Miville-Desch{\^e}nes}, {Molinari}, {Montier}, {Morgante}, {Moss}, {Natoli},
  {N{\o}rgaard-Nielsen}, {Pagano}, {Paoletti}, {Partridge}, {Patanchon},
  {Peiris}, {Perrotta}, {Pettorino}, {Piacentini}, {Polastri}, {Polenta},
  {Puget}, {Rachen}, {Reinecke}, {Remazeilles}, {Renzi}, {Rocha}, {Rosset},
  {Roudier}, {Rubi{\~n}o-Mart{\'{\i}}n}, {Ruiz-Granados}, {Salvati}, {Sandri},
  {Savelainen}, {Scott}, {Shellard}, {Sirignano}, {Sirri}, {Spencer},
  {Sunyaev}, {Suur-Uski}, {Tauber}, {Tavagnacco}, {Tenti}, {Toffolatti},
  {Tomasi}, {Trombetti}, {Valenziano}, {Valiviita}, {Van Tent}, {Vibert},
  {Vielva}, {Villa}, {Vittorio}, {Wandelt}, {Wehus}, {White}, {White},
  {Zacchei}, \& {Zonca}}]{Planck18}
{Planck Collaboration}, {Aghanim}, N., {Akrami}, Y., {et~al.} 2018, arXiv
  e-prints, arXiv:1807.06209

\bibitem[{{Porciani} \& {Norberg}(2006)}]{Porciani06}
{Porciani}, C., \& {Norberg}, P. 2006, \mnras, 371, 1824

\bibitem[{{Richards} {et~al.}(2002){Richards}, {Fan}, {Newberg}, {Strauss},
  {Vanden Berk}, {Schneider}, {Yanny}, {Boucher}, {Burles}, {Frieman}, {Gunn},
  {Hall}, {Ivezi{\'c}}, {Kent}, {Loveday}, {Lupton}, {Rockosi}, {Schlegel},
  {Stoughton}, {SubbaRao}, \& {York}}]{Richards02b}
{Richards}, G.~T., {Fan}, X., {Newberg}, H.~J., {et~al.} 2002, \aj, 123, 2945

\bibitem[{{Schlegel} {et~al.}(1998){Schlegel}, {Finkbeiner}, \&
  {Davis}}]{Schlegel98}
{Schlegel}, D.~J., {Finkbeiner}, D.~P., \& {Davis}, M. 1998, \apj, 500, 525

\bibitem[{{Shanks} {et~al.}(2015){Shanks}, {Metcalfe}, {Chehade}, {Findlay},
  {Irwin}, {Gonzalez-Solares}, {Lewis}, {Yoldas}, {Mann}, {Read}, {Sutorius},
  \& {Voutsinas}}]{Shanks15}
{Shanks}, T., {Metcalfe}, N., {Chehade}, B., {et~al.} 2015, \mnras, 451, 4238

\bibitem[{{Shen} {et~al.}(2007){Shen}, {Strauss}, {Oguri}, {Hennawi}, {Fan},
  {Richards}, {Hall}, {Gunn}, {Schneider}, {Szalay}, {Thakar}, {Vanden Berk},
  {Anderson}, {Bahcall}, {Connolly}, \& {Knapp}}]{Shen07}
{Shen}, Y., {Strauss}, M.~A., {Oguri}, M., {et~al.} 2007, \aj, 133, 2222

\bibitem[{{Shen} {et~al.}(2009){Shen}, {Strauss}, {Ross}, {Hall}, {Lin},
  {Richards}, {Schneider}, {Weinberg}, {Connolly}, {Fan}, {Hennawi}, {Shankar},
  {Vanden Berk}, {Bahcall}, \& {Brunner}}]{Shen09}
{Shen}, Y., {Strauss}, M.~A., {Ross}, N.~P., {et~al.} 2009, \apj, 697, 1656

\bibitem[{{Shen} {et~al.}(2013){Shen}, {McBride}, {White}, {Zheng}, {Myers},
  {Guo}, {Kirkpatrick}, {Padmanabhan}, {Parejko}, {Ross}, {Schlegel},
  {Schneider}, {Streblyanska}, {Swanson}, {Zehavi}, {Pan}, {Bizyaev},
  {Brewington}, {Ebelke}, {Malanushenko}, {Malanushenko}, {Oravetz}, {Simmons},
  \& {Snedden}}]{Shen13}
{Shen}, Y., {McBride}, C.~K., {White}, M., {et~al.} 2013, \apj, 778, 98

\bibitem[{{Shen} {et~al.}(2016){Shen}, {Brandt}, {Richards}, {Denney},
  {Greene}, {Grier}, {Ho}, {Peterson}, {Petitjean}, {Schneider}, {Tao}, \&
  {Trump}}]{Shen16}
{Shen}, Y., {Brandt}, W.~N., {Richards}, G.~T., {et~al.} 2016, \apj, 831, 7

\bibitem[{{Shibuya} {et~al.}(2019){Shibuya}, {Ouchi}, {Harikane}, \&
  {Nakajima}}]{Shibuya19}
{Shibuya}, T., {Ouchi}, M., {Harikane}, Y., \& {Nakajima}, K. 2019, \apj, 871,
  164

\bibitem[{{Shimasaku} {et~al.}(2006){Shimasaku}, {Kashikawa}, {Doi}, {Ly},
  {Malkan}, {Matsuda}, {Ouchi}, {Hayashino}, {Iye}, {Motohara}, {Murayama},
  {Nagao}, {Ohta}, {Okamura}, {Sasaki}, {Shioya}, \& {Taniguchi}}]{Shimasaku06}
{Shimasaku}, K., {Kashikawa}, N., {Doi}, M., {et~al.} 2006, \pasj, 58, 313

\bibitem[{{Simpson} {et~al.}(2014){Simpson}, {Mortlock}, {Warren}, {Cantalupo},
  {Hewett}, {McLure}, {McMahon}, \& {Venemans}}]{Simpson14}
{Simpson}, C., {Mortlock}, D., {Warren}, S., {et~al.} 2014, \mnras, 442, 3454

\bibitem[{{Sobral} {et~al.}(2018){Sobral}, {Santos}, {Matthee},
  {Paulino-Afonso}, {Ribeiro}, {Calhau}, \& {Khostovan}}]{Sobral18}
{Sobral}, D., {Santos}, S., {Matthee}, J., {et~al.} 2018, \mnras, 476, 4725

\bibitem[{{Springel} {et~al.}(2005){Springel}, {Di Matteo}, \&
  {Hernquist}}]{springel05}
{Springel}, V., {Di Matteo}, T., \& {Hernquist}, L. 2005, \mnras, 361, 776

\bibitem[{{Stetson}(2000)}]{Stetson00}
{Stetson}, P.~B. 2000, \pasp, 112, 925

\bibitem[{{Stiavelli} {et~al.}(2005){Stiavelli}, {Djorgovski}, {Pavlovsky},
  {Scarlata}, {Stern}, {Mahabal}, {Thompson}, {Dickinson}, {Panagia}, \&
  {Meylan}}]{Stiavelli05}
{Stiavelli}, M., {Djorgovski}, S.~G., {Pavlovsky}, C., {et~al.} 2005, \apjl,
  622, L1

\bibitem[{{Swinbank} {et~al.}(2012){Swinbank}, {Baker}, {Barr}, {Hook}, \&
  {Bland-Hawthorn}}]{Swinbank12}
{Swinbank}, J., {Baker}, J., {Barr}, J., {Hook}, I., \& {Bland-Hawthorn}, J.
  2012, \mnras, 422, 2980

\bibitem[{{Trainor} \& {Steidel}(2012)}]{Trainor12}
{Trainor}, R.~F., \& {Steidel}, C.~C. 2012, \apj, 752, 39

\bibitem[{{Trakhtenbrot} {et~al.}(2017){Trakhtenbrot}, {Lira}, {Netzer},
  {Cicone}, {Maiolino}, \& {Shemmer}}]{Trakhtenbrot17}
{Trakhtenbrot}, B., {Lira}, P., {Netzer}, H., {et~al.} 2017, \apj, 836, 8

\bibitem[{{Tytler} \& {Fan}(1992)}]{Tytler92}
{Tytler}, D., \& {Fan}, X.-M. 1992, \apjs, 79, 1

\bibitem[{{Uchiyama} {et~al.}(2018){Uchiyama}, {Toshikawa}, {Kashikawa},
  {Overzier}, {Chiang}, {Marinello}, {Tanaka}, {Niino}, {Ishikawa}, {Onoue},
  {Ichikawa}, {Akiyama}, {Coupon}, {Harikane}, {Imanishi}, {Kodama},
  {Komiyama}, {Lee}, {Lin}, {Miyazaki}, {Nagao}, {Nishizawa}, {Ono}, {Ouchi},
  \& {Wang}}]{Uchiyama18}
{Uchiyama}, H., {Toshikawa}, J., {Kashikawa}, N., {et~al.} 2018, \pasj, 70, S32

\bibitem[{{Utsumi} {et~al.}(2010){Utsumi}, {Goto}, {Kashikawa}, {Miyazaki},
  {Komiyama}, {Furusawa}, \& {Overzier}}]{Utsumi10}
{Utsumi}, Y., {Goto}, T., {Kashikawa}, N., {et~al.} 2010, \apj, 721, 1680

\bibitem[{{Vanden Berk} {et~al.}(2001){Vanden Berk}, {Richards}, {Bauer},
  {Strauss}, {Schneider}, {Heckman}, {York}, {Hall}, {Fan}, {Knapp},
  {Anderson}, {Annis}, {Bahcall}, {Bernardi}, {Briggs}, {Brinkmann}, {Brunner},
  {Burles}, {Carey}, {Castander}, {Connolly}, {Crocker}, {Csabai}, {Doi},
  {Finkbeiner}, {Friedman}, {Frieman}, {Fukugita}, {Gunn}, {Hennessy},
  {Ivezi{\'c}}, {Kent}, {Kunszt}, {Lamb}, {Leger}, {Long}, {Loveday}, {Lupton},
  {Meiksin}, {Merelli}, {Munn}, {Newberg}, {Newcomb}, {Nichol}, {Owen}, {Pier},
  {Pope}, {Rockosi}, {Schlegel}, {Siegmund}, {Smee}, {Snir}, {Stoughton},
  {Stubbs}, {SubbaRao}, {Szalay}, {Szokoly}, {Tremonti}, {Uomoto}, {Waddell},
  {Yanny}, \& {Zheng}}]{VandenBerk01}
{Vanden Berk}, D.~E., {Richards}, G.~T., {Bauer}, A., {et~al.} 2001, \aj, 122,
  549

\bibitem[{{Venemans} {et~al.}(2005){Venemans}, {R{\"o}ttgering}, {Miley},
  {Kurk}, {De Breuck}, {Overzier}, {van Breugel}, {Carilli}, {Ford}, {Heckman},
  {Pentericci}, \& {McCarthy}}]{Venemans05}
{Venemans}, B.~P., {R{\"o}ttgering}, H.~J.~A., {Miley}, G.~K., {et~al.} 2005,
  \aap, 431, 793

\bibitem[{{Venemans} {et~al.}(2007){Venemans}, {R{\"o}ttgering}, {Miley}, {van
  Breugel}, {de Breuck}, {Kurk}, {Pentericci}, {Stanford}, {Overzier}, {Croft},
  \& {Ford}}]{Venemans07}
---. 2007, \aap, 461, 823

\bibitem[{{White} {et~al.}(2008){White}, {Martini}, \& {Cohn}}]{White08}
{White}, M., {Martini}, P., \& {Cohn}, J.~D. 2008, \mnras, 390, 1179

\bibitem[{{White} {et~al.}(2012){White}, {Myers}, {Ross}, {Schlegel},
  {Hennawi}, {Shen}, {McGreer}, {Strauss}, {Bolton}, {Bovy}, {Fan},
  {Miralda-Escude}, {Palanque-Delabrouille}, {Paris}, {Petitjean}, {Schneider},
  {Viel}, {Weinberg}, {Yeche}, {Zehavi}, {Pan}, {Snedden}, {Bizyaev},
  {Brewington}, {Brinkmann}, {Malanushenko}, {Malanushenko}, {Oravetz},
  {Simmons}, {Sheldon}, \& {Weaver}}]{White12}
{White}, M., {Myers}, A.~D., {Ross}, N.~P., {et~al.} 2012, \mnras, 424, 933

\bibitem[{{Willott} {et~al.}(2005){Willott}, {Percival}, {McLure}, {Crampton},
  {Hutchings}, {Jarvis}, {Sawicki}, \& {Simard}}]{Willott05}
{Willott}, C.~J., {Percival}, W.~J., {McLure}, R.~J., {et~al.} 2005, \apj, 626,
  657

\bibitem[{{Worseck} \& {Prochaska}(2011)}]{Worseck11}
{Worseck}, G., \& {Prochaska}, J.~X. 2011, \apj, 728, 23

\bibitem[{{York} {et~al.}(2000){York}, {Adelman}, {Anderson}, {Anderson},
  {Annis}, {Bahcall}, {Bakken}, {Barkhouser}, {Bastian}, {Berman}, {Boroski},
  {Bracker}, {Briegel}, {Briggs}, {Brinkmann}, {Brunner}, {Burles}, {Carey},
  {Carr}, {Castander}, {Chen}, {Colestock}, {Connolly}, {Crocker}, {Csabai},
  {Czarapata}, {Davis}, {Doi}, {Dombeck}, {Eisenstein}, {Ellman}, {Elms},
  {Evans}, {Fan}, {Federwitz}, {Fiscelli}, {Friedman}, {Frieman}, {Fukugita},
  {Gillespie}, {Gunn}, {Gurbani}, {de Haas}, {Haldeman}, {Harris}, {Hayes},
  {Heckman}, {Hennessy}, {Hindsley}, {Holm}, {Holmgren}, {Huang}, {Hull},
  {Husby}, {Ichikawa}, {Ichikawa}, {Ivezi{\'c}}, {Kent}, {Kim}, {Kinney},
  {Klaene}, {Kleinman}, {Kleinman}, {Knapp}, {Korienek}, {Kron}, {Kunszt},
  {Lamb}, {Lee}, {Leger}, {Limmongkol}, {Lindenmeyer}, {Long}, {Loomis},
  {Loveday}, {Lucinio}, {Lupton}, {MacKinnon}, {Mannery}, {Mantsch}, {Margon},
  {McGehee}, {McKay}, {Meiksin}, {Merelli}, {Monet}, {Munn}, {Narayanan},
  {Nash}, {Neilsen}, {Neswold}, {Newberg}, {Nichol}, {Nicinski}, {Nonino},
  {Okada}, {Okamura}, {Ostriker}, {Owen}, {Pauls}, {Peoples}, {Peterson},
  {Petravick}, {Pier}, {Pope}, {Pordes}, {Prosapio}, {Rechenmacher}, {Quinn},
  {Richards}, {Richmond}, {Rivetta}, {Rockosi}, {Ruthmansdorfer}, {Sandford},
  {Schlegel}, {Schneider}, {Sekiguchi}, {Sergey}, {Shimasaku}, {Siegmund},
  {Smee}, {Smith}, {Snedden}, {Stone}, {Stoughton}, {Strauss}, {Stubbs},
  {SubbaRao}, {Szalay}, {Szapudi}, {Szokoly}, {Thakar}, {Tremonti}, {Tucker},
  {Uomoto}, {Vanden Berk}, {Vogeley}, {Waddell}, {Wang}, {Watanabe},
  {Weinberg}, {Yanny}, {Yasuda}, \& {SDSS Collaboration}}]{York00}
{York}, D.~G., {Adelman}, J., {Anderson}, Jr., J.~E., {et~al.} 2000, \aj, 120,
  1579

\bibitem[{{Zheng} {et~al.}(2006){Zheng}, {Overzier}, {Bouwens}, {White},
  {Ford}, {Ben{\'{\i}}tez}, {Blakeslee}, {Bradley}, {Jee}, {Martel}, {Mei},
  {Zirm}, {Illingworth}, {Clampin}, {Hartig}, {Ardila}, {Bartko}, {Broadhurst},
  {Brown}, {Burrows}, {Cheng}, {Cross}, {Demarco}, {Feldman}, {Franx},
  {Golimowski}, {Goto}, {Gronwall}, {Holden}, {Homeier}, {Infante}, {Kimble},
  {Krist}, {Lesser}, {Menanteau}, {Meurer}, {Miley}, {Motta}, {Postman},
  {Rosati}, {Sirianni}, {Sparks}, {Tran}, \& {Tsvetanov}}]{Zheng06}
{Zheng}, W., {Overzier}, R.~A., {Bouwens}, R.~J., {et~al.} 2006, \apj, 640, 574

\end{thebibliography}

\end{document}